%% file: Multiplicity-main.tex
\pdfoutput=1	

\documentclass[12pt,a4paper]{article}

\usepackage{ifthen} 
\newboolean{pdflatex}	
\setboolean{pdflatex}{true} 

\newboolean{articletitles}
\setboolean{articletitles}{true} 

\newboolean{uprightparticles}
\setboolean{uprightparticles}{false} 

\newboolean{inbibliography}
\setboolean{inbibliography}{false} 

\input{preamble}
\usepackage{subfigure}
\begin{document}

\renewcommand{\thefootnote}{\fnsymbol{footnote}}
\setcounter{footnote}{1}

\input{title-LHCb-PAPER}


\renewcommand{\thefootnote}{\arabic{footnote}}
\setcounter{footnote}{0}



\pagestyle{plain} 
\setcounter{page}{1}
\pagenumbering{arabic}

\input{introduction}

\input{detector}

\input{dataset}

\input{eventdefinition}

\input{analysis}

\input{systematics}

\input{results}

\input{summary}

\input{acknowledgements}

\input{appendix}

\addcontentsline{toc}{section}{References}
\setboolean{inbibliography}{true}
\bibliographystyle{LHCb}
\bibliography{Multiplicity-main,LHCb-PAPER,LHCb-CONF,LHCb-DP}

\end{document}

%% file: preamble.tex
\usepackage{microtype}
\usepackage{lineno}  
\usepackage{xspace} 

\usepackage{graphicx}  
\usepackage{color}
\usepackage{colortbl}
\graphicspath{{./figs/}} 

\usepackage{amsmath} 
\usepackage{amssymb}
\usepackage{amsfonts}
\usepackage{upgreek} 

\newcommand*\patchAmsMathEnvironmentForLineno[1]{%
\expandafter\let\csname old#1\expandafter\endcsname\csname #1\endcsname
\expandafter\let\csname oldend#1\expandafter\endcsname\csname
end#1\endcsname
 \renewenvironment{#1}%
   {\linenomath\csname old#1\endcsname}%
   {\csname oldend#1\endcsname\endlinenomath}%
}
\newcommand*\patchBothAmsMathEnvironmentsForLineno[1]{%
  \patchAmsMathEnvironmentForLineno{#1}%
  \patchAmsMathEnvironmentForLineno{#1*}%
}
\AtBeginDocument{%
\patchBothAmsMathEnvironmentsForLineno{equation}%
\patchBothAmsMathEnvironmentsForLineno{align}%
\patchBothAmsMathEnvironmentsForLineno{flalign}%
\patchBothAmsMathEnvironmentsForLineno{alignat}%
\patchBothAmsMathEnvironmentsForLineno{gather}%
\patchBothAmsMathEnvironmentsForLineno{multline}%
}

\usepackage{hyperref}    
\usepackage[all]{hypcap} 

\input{lhcb-symbols-def} 

\usepackage{cite} 
\usepackage{mciteplus}

%% file: lhcb-symbols-def.tex
\def\lhcb {\mbox{LHCb}\xspace}
\def\ux85 {\mbox{UX85}\xspace}

\def\lhc    {\mbox{LHC}\xspace}

\def\velo   {VELO\xspace}

\ifthenelse{\boolean{uprightparticles}}%
{

 \def\PDelta      {\ensuremath{\Delta}\xspace}                 
 \def\PXi      {\ensuremath{\Xi}\xspace}                 
 \def\PLambda      {\ensuremath{\Lambda}\xspace}                 
 \def\PSigma      {\ensuremath{\Sigma}\xspace}                 
 \def\POmega      {\ensuremath{\Omega}\xspace}                 
 \def\PUpsilon      {\ensuremath{\Upsilon}\xspace}                 
 
 \def\PB      {\ensuremath{\mathrm{B}}\xspace}                 
                  
 \def\PD      {\ensuremath{\mathrm{D}}\xspace}

 \def\PK      {\ensuremath{\mathrm{K}}\xspace}

 \def\Pb      {\ensuremath{\mathrm{b}}\xspace}                 
 \def\Pc      {\ensuremath{\mathrm{c}}\xspace}

 \def\Pi      {\ensuremath{\mathrm{i}}\xspace}

}
{

 \mathchardef\PDelta="7101
 \mathchardef\PXi="7104
 \mathchardef\PLambda="7103
 \mathchardef\PSigma="7106
 \mathchardef\POmega="710A
 \mathchardef\PUpsilon="7107
                  
 \def\PB      {\ensuremath{B}\xspace}                 
                  
 \def\PD      {\ensuremath{D}\xspace}

 \def\PK      {\ensuremath{K}\xspace}

 \def\Pb      {\ensuremath{b}\xspace}                 
 \def\Pc      {\ensuremath{c}\xspace}

 \def\Pi      {\ensuremath{i}\xspace}

}

\def\cquark    {\ensuremath{\Pc}\xspace}

\def\bquark    {\ensuremath{\Pb}\xspace}

\def\kaon  {\ensuremath{\PK}\xspace}
  \def\Kbar  {\kern 0.2em\overline{\kern -0.2em \PK}{}\xspace}

\def\Kz    {\ensuremath{\kaon^0}\xspace}
\def\Kzb   {\ensuremath{\Kbar^0}\xspace}
\def\KzKzb {\ensuremath{\Kz \kern -0.16em \Kzb}\xspace}
\def\Kp    {\ensuremath{\kaon^+}\xspace}
\def\Km    {\ensuremath{\kaon^-}\xspace}

\def\KpKm  {\ensuremath{\Kp \kern -0.16em \Km}\xspace}

  \def\Dbar    {\kern 0.2em\overline{\kern -0.2em \PD}{}\xspace}
\def\D       {\ensuremath{\PD}\xspace}

\def\Dz      {\ensuremath{\D^0}\xspace}
\def\Dzb     {\ensuremath{\Dbar^0}\xspace}
\def\DzDzb   {\ensuremath{\Dz {\kern -0.16em \Dzb}}\xspace}
\def\Dp      {\ensuremath{\D^+}\xspace}
\def\Dm      {\ensuremath{\D^-}\xspace}

\def\DpDm    {\ensuremath{\Dp {\kern -0.16em \Dm}}\xspace}

\def\Bbar    {\ensuremath{\kern 0.18em\overline{\kern -0.18em \PB}{}}\xspace}

  \def\Y#1S{\ensuremath{\PUpsilon{(#1S)}}\xspace}

\def\Lbar {\ensuremath{\kern 0.1em\overline{\kern -0.1em\PLambda}}\xspace}

\def\AT#1     {\ensuremath{A_{\mathrm{T}}^{#1}}\xspace}           

\def\C#1      {\ensuremath{\mathcal{C}_{#1}}\xspace}                       
\def\Cp#1     {\ensuremath{\mathcal{C}_{#1}^{'}}\xspace}                    
\def\Ceff#1   {\ensuremath{\mathcal{C}_{#1}^{\mathrm{(eff)}}}\xspace}        
\def\Cpeff#1  {\ensuremath{\mathcal{C}_{#1}^{'\mathrm{(eff)}}}\xspace}       
\def\Ope#1    {\ensuremath{\mathcal{O}_{#1}}\xspace}                       
\def\Opep#1   {\ensuremath{\mathcal{O}_{#1}^{'}}\xspace}                    

\newcommand{\tev}{\ifthenelse{\boolean{inbibliography}}{\ensuremath{~T\kern -0.05em eV}\xspace}{\ensuremath{\mathrm{\,Te\kern -0.1em V}}\xspace}}
\newcommand{\gev}{\ensuremath{\mathrm{\,Ge\kern -0.1em V}}\xspace}
\newcommand{\mev}{\ensuremath{\mathrm{\,Me\kern -0.1em V}}\xspace}
\newcommand{\kev}{\ensuremath{\mathrm{\,ke\kern -0.1em V}}\xspace}
\newcommand{\ev}{\ensuremath{\mathrm{\,e\kern -0.1em V}}\xspace}
\newcommand{\gevc}{\ensuremath{{\mathrm{\,Ge\kern -0.1em V\!/}c}}\xspace}
\newcommand{\mevc}{\ensuremath{{\mathrm{\,Me\kern -0.1em V\!/}c}}\xspace}
\newcommand{\gevcc}{\ensuremath{{\mathrm{\,Ge\kern -0.1em V\!/}c^2}}\xspace}
\newcommand{\gevgevcccc}{\ensuremath{{\mathrm{\,Ge\kern -0.1em V^2\!/}c^4}}\xspace}
\newcommand{\mevcc}{\ensuremath{{\mathrm{\,Me\kern -0.1em V\!/}c^2}}\xspace}

\def\mm   {\ensuremath{\rm \,mm}\xspace}

\def\mum  {\ensuremath{\,\upmu\rm m}\xspace}

\def\ps   {\ensuremath{{\rm \,ps}}\xspace}

\def\gsim{{~\raise.15em\hbox{$>$}\kern-.85em
          \lower.35em\hbox{$\sim$}~}\xspace}
\def\lsim{{~\raise.15em\hbox{$<$}\kern-.85em
          \lower.35em\hbox{$\sim$}~}\xspace}

\def\sqs   {\ensuremath{\protect\sqrt{s}}\xspace}

\def\ptot       {\mbox{$p$}\xspace}
\def\pt         {\mbox{$p_{\rm T}$}\xspace}

\def\evtgen     {\mbox{\textsc{EvtGen}}\xspace}

\def\geant      {\mbox{\textsc{Geant4}}\xspace}

\def\herwig     {\mbox{\textsc{Herwig}}\xspace}

\def\photos     {\mbox{\textsc{Photos}}\xspace}

\def\pythia     {\mbox{\textsc{Pythia}}\xspace}

\def\tell1  {TELL1\xspace}
\def\ukl1   {UKL1\xspace}

\def\phojet     {\mbox{\textsc{Phojet}}\xspace}
\def\herwig     {\mbox{\textsc{Herwig++}}\xspace}

%% file: title-LHCb-PAPER.tex

\begin{titlepage}
\pagenumbering{roman}

\vspace*{-1.5cm}
\centerline{\large EUROPEAN ORGANIZATION FOR NUCLEAR RESEARCH (CERN)}
\vspace*{1.5cm}
\hspace*{-0.5cm}
\begin{tabular*}{\linewidth}{lc@{\extracolsep{\fill}}r}
\ifthenelse{\boolean{pdflatex}}
{\vspace*{-2.7cm}\mbox{\!\!\!\includegraphics[width=.14\textwidth]{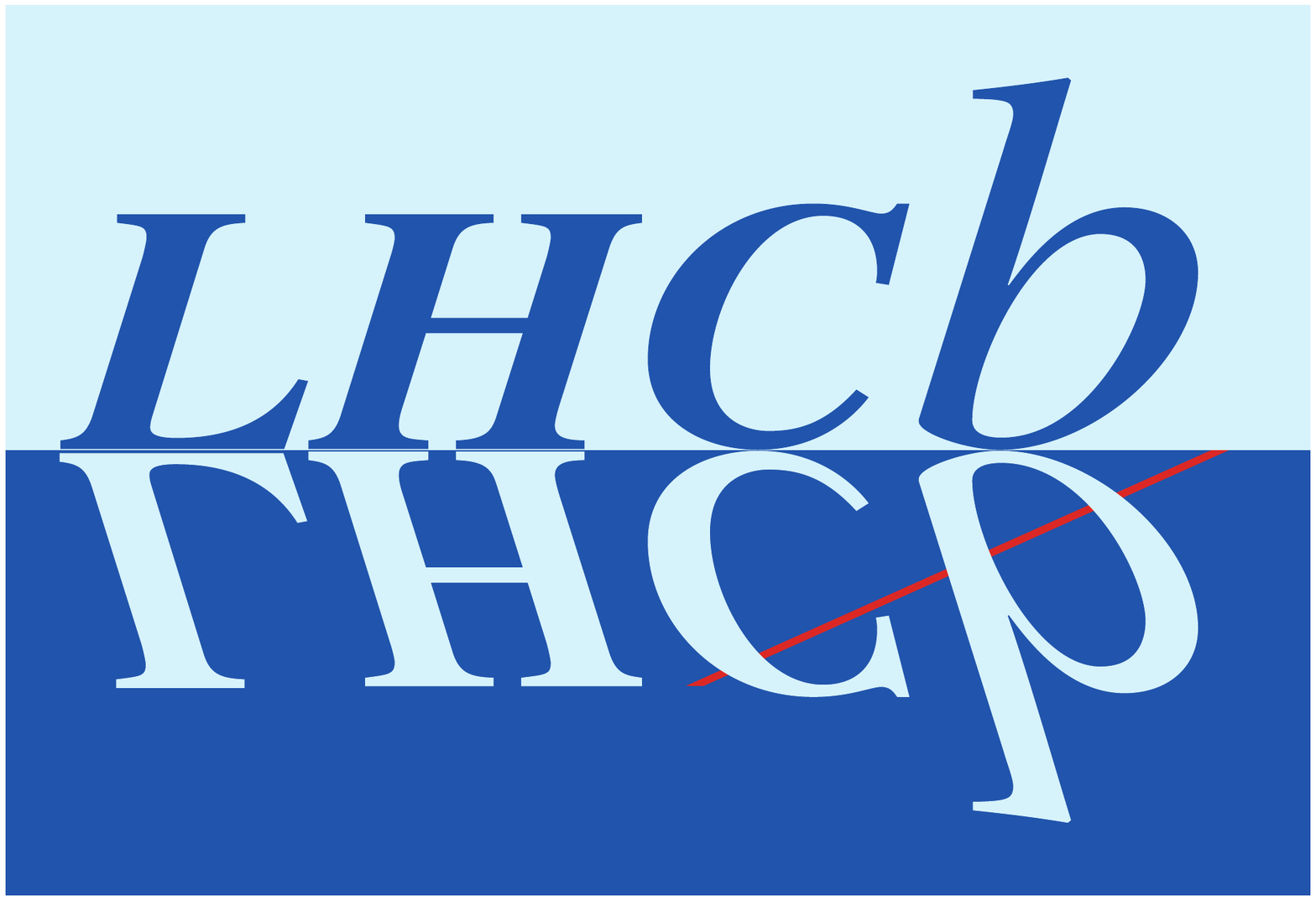}} & &}%
{\vspace*{-1.2cm}\mbox{\!\!\!\includegraphics[width=.12\textwidth]{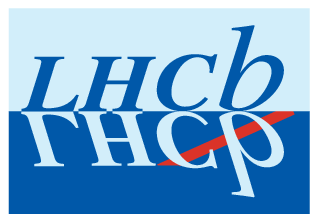}} & &}%
\\
 & & CERN-PH-EP-2014-023 \\  
 & & LHCb-PAPER-2013-070 \\  
 & & April 30, 2014 \\ 
\end{tabular*}

\vspace*{2.5cm}

{\bf\boldmath\huge
\begin{center}
  Measurement of charged particle multiplicities and densities in $pp$ collisions at $\sqrt{s}=7\;$TeV in the forward region
\end{center}
}

\vspace*{1.2cm}

\begin{center}
The LHCb collaboration\footnote{Authors are listed on the following pages.}
\end{center}

\vspace{\fill}

\begin{abstract}
  \noindent
  Charged particle multiplicities are studied in proton-proton collisions in the forward region at a centre-of-mass energy of $\sqrt{s} = 7\;$TeV with data collected by the LHCb detector. The forward spectrometer allows access to a kinematic range of $2.0<\eta<4.8$ in pseudorapidity, momenta greater than $2\;\mbox{GeV/}c$ and transverse momenta greater than $0.2\;\mbox{GeV/}c$. The measurements are performed using events with at least one charged particle in the kinematic acceptance. The results are presented as functions of pseudorapidity and transverse momentum and are compared to predictions from several Monte Carlo event generators.
\end{abstract}

\vspace*{1.2cm}

\begin{center}
  Submitted to European Physical Journal C
  \end{center}

\vspace{\fill}

{\footnotesize 
\centerline{\copyright~CERN on behalf of the \lhcb collaboration, license \href{http://creativecommons.org/licenses/by/3.0/}{CC-BY-3.0}.}}
\vspace*{2mm}

\end{titlepage}


\newpage
\setcounter{page}{2}
\mbox{~}
\newpage

\input{LHCb_HD_authorlist_2013-12-18.tex}

\cleardoublepage

%% file: LHCb_HD_authorlist_2013-12-18.tex
\centerline{\large\bf LHCb collaboration}
\begin{flushleft}
\small
R.~Aaij$^{41}$, 
B.~Adeva$^{37}$, 
M.~Adinolfi$^{46}$, 
A.~Affolder$^{52}$, 
Z.~Ajaltouni$^{5}$, 
J.~Albrecht$^{9}$, 
F.~Alessio$^{38}$, 
M.~Alexander$^{51}$, 
S.~Ali$^{41}$, 
G.~Alkhazov$^{30}$
% ...

{\footnotesize \it
% ...
$ ^{5}$Clermont Universit\'{e}, Universit\'{e} Blaise Pascal, CNRS/IN2P3, LPC, Clermont-Ferrand, France\\
% ...
$ ^{9}$Fakult\"{a}t Physik, Technische Universit\"{a}t Dortmund, Dortmund, Germany\\
% ...
$ ^{30}$Petersburg Nuclear Physics Institute (PNPI), Gatchina, Russia\\
% ...
$ ^{37}$Universidad de Santiago de Compostela, Santiago de Compostela, Spain\\
$ ^{38}$European Organization for Nuclear Research (CERN), Geneva, Switzerland\\
% ...
$ ^{41}$Nikhef National Institute for Subatomic Physics, Amsterdam, The Netherlands\\
% ...
$ ^{46}$H.H. Wills Physics Laboratory, University of Bristol, Bristol, United Kingdom\\
% ...
$ ^{51}$School of Physics and Astronomy, University of Glasgow, Glasgow, United Kingdom\\
$ ^{52}$Oliver Lodge Laboratory, University of Liverpool, Liverpool, United Kingdom\\
% ...
}
\end{flushleft}

%% file: introduction.tex

\section{Introduction}
\label{sec:Introduction}

The phenomenology of soft quantum chromodynamic (QCD) processes such as light particle production in proton-proton ($pp$) collisions cannot be predicted using perturbative calculations, but can be described by models implemented in Monte Carlo event generators. 
The calculation of the fragmentation and hadronization processes as well as the modelling of the final states~\cite{Kaidalov:1983ew, Capella1994225} arising from the soft component of a collision (underlying event) are treated differently in the various event generators. 
The phenomenological models contain parameters that need to be tuned depending on the collision energy and colliding particles species. 
This is typically achieved using soft QCD measurements. 
The \lhcb collaboration reported measurements on energy flow~\cite{LHCb-PAPER-2012-034}, production cross-sections~\cite{LHCb-PAPER-2010-001, LHCb-PAPER-2011-007} and production ratios of various particle species~\cite{LHCb-PAPER-2011-037} in the forward region, all of which provide information for event generator optimization. 

A fundamental input used for the tuning process is the measurement of prompt charged particle multiplicities. 
In combination with the study of the corresponding momentum spectra and angular distributions, these measurements can be used to gain a better understanding of hadron collisions. 
An accurate description of the underlying event is vital for understanding backgrounds in beyond the Standard Model searches or precision measurements of the Standard Model parameters. 
Previous measurements of charged particle multiplicities performed with $pp$ collisions at the Large Hadron Collider (LHC) were reported by the ATLAS~\cite{Aad:2010rd, ATLAS:2010ir}, CMS~\cite{Khachatryan:2010nk} and ALICE~\cite{Aamodt:2010ft,Aamodt:2010pp} collaborations. 
All of these measurements were performed in the central pseudorapidity region. The forward region was studied with the LHCb detector, where an inclusive multiplicity measurement without momentum information was performed~\cite{LHCb-PAPER-2011-011}. 

In this paper, $pp$ interactions at a centre-of-mass energy of $\sqrt{s}=7\;$TeV that produce at least one prompt charged particle in the pseudorapidity range of $2.0<\eta<4.8$, with a momentum of $\ptot > 2\gevc$ and transverse momentum of $\pt > 0.2\gevc$, are studied. 
A prompt particle is defined as a particle that either originates directly from the primary vertex or from a decay chain in which the sum of mean lifetimes does not exceed $10\ps$. 
As a consequence, decay products of beauty and charm hadrons are treated as prompt particles. 
The information from the full tracking system of the \lhcb detector is used, which permits the measurement of the momentum dependence of charged particle multiplicities. 
Multiplicity distributions, $P(n)$, for prompt charged particles are reported for the total accessible phase space region as well as for $\eta$ and \pt ranges. 
In addition, mean particle densities are presented as functions of transverse momentum, $dn/d\pt$, and of pseudorapidity, $dn/d\eta$.

The paper is organised as follows. In Sect.~\ref{sec:Detector} a brief description of the \lhcb detector and an overview of track reconstruction algorithms are provided. The recorded data set and Monte Carlo simulations are described in Sect.~\ref{sec:Dataset}, followed by a discussion of the definition of visible event and the data selection in Sect.~\ref{sec:Eventdefinition}. The analysis method is described in Sect.~\ref{sec:AnalysisStrategy}, and systematic uncertainties are given in Sect.~\ref{sec:Systematics}. The final results are compared to event generator predictions in Sects.~\ref{sec:ResultsDensity} and \ref{sec:ResultsMultiplicity}, before summarising in Sect.~\ref{sec:Summary}. 

%% file: detector.tex

\section{LHCb detector and track reconstruction}
\label{sec:Detector}
The \lhcb detector~\cite{Alves:2008zz} is a single-arm forward spectrometer covering the \mbox{pseudorapidity} range $2<\eta <5$, designed for the study of particles containing \bquark or \cquark quarks. 
The detector includes a high-precision tracking system consisting of a silicon-strip vertex detector (VELO) surrounding the $pp$ interaction region, a large-area silicon-strip detector located upstream of a dipole magnet with a bending power of about $4{\rm\,Tm}$, and three stations of silicon-strip detectors and straw drift tubes placed downstream.
The combined tracking system provides a momentum measurement with relative uncertainty that varies from 0.4\,\% at 2\gevc to 0.6\,\% at 100\gevc, and impact parameter resolution of 20\mum for tracks with large transverse momentum. 
The direction of the magnetic field of the spectrometer dipole magnet is reversed regularly. 
Different types of charged hadrons are distinguished by information from two ring-imaging Cherenkov detectors. 
Photon, electron and hadron candidates are identified by a calorimeter system consisting of scintillating-pad and preshower detectors, an electromagnetic calorimeter and a hadronic calorimeter. 
Muons are identified by a system composed of alternating layers of iron and multiwire proportional chambers. 
The trigger consists of a hardware stage, based on information from the calorimeter and muon systems, followed by a software stage, which applies full event reconstruction.

The reconstruction algorithms provide different track types depending on the sub-detectors considered. 
Only two types of tracks are used in this analysis. 
\velo tracks are only reconstructed in the \velo sub-detector and provide no momentum information. 
\textit{Long} tracks are reconstructed by extrapolating \velo tracks through the magnetic dipole field and matching them with hits in the downstream tracking stations, providing momentum information. 
This is the highest-quality track type and is used for most physics analyses. 
Requiring charged particles to stay within the geometric acceptance of the \lhcb detector after deflection by the magnetic field further restricts the accessible phase space to a minimum momentum of around $2\gevc$. 
The \lhcb detector design minimizes the material of the tracking detectors and allows a high track-reconstruction efficiency even for particles with low momenta. 
However, the limited number of tracking stations results in the presence of misreconstructed (\textit{fake}) tracks. 
A reconstructed track is considered as fake if it does not correspond to the trajectory of a genuine charged particle. 
The fraction of fake long tracks is non-negligible as the extrapolation of a track through the magnetic field is performed over a distance of several metres, resulting in wrong association between \velo tracks and track segments reconstructed downstream. 
Another source of wrong track assignment arises from duplicate tracks. 
These track pairs either share a certain number of hits or consist of different track segments originating from a single particle. 

%% file: dataset.tex

\section{Data set and simulation}
\label{sec:Dataset}

The measurements are performed using a minimum-bias data sample of $pp$ collisions at a centre-of-mass energy of \sqs=7\tev collected during 2010. 
In this low-luminosity running period, the average number of interactions in the detector acceptance per recorded bunch crossing was less than $0.1$. 
The contribution from bunch crossings with more than one collision (\textit{pile-up} events) is determined to be less than $4\,\%$ and is considered as a correction in the analysis. 
The data consists of 3 million events recorded in equal proportion for both magnetic field polarities. 
The low luminosity and interaction rate of the proton beams allowed the \lhcb detector to be operated with a simplified trigger scheme. 
For the minimum-bias data set of this analysis, the hardware stage of the trigger system accepted all events, which were then reconstructed by the higher-level software trigger. 
Events with at least one reconstructed track segment in the VELO were selected.

Fully simulated minimum-bias $pp$ collisions are generated using the \pythia 6.4 event generator~\cite{Sjostrand:2006za} with a specific \lhcb configuration~\cite{LHCb-PROC-2010-056} using CTEQ6L~\cite{Pumplin:2002vw} parton density functions (PDFs). 
This implementation, called the \lhcb tune, contains contributions from elastic and inelastic processes, where the latter also include single and double diffractive components. 
Decays of hadrons are performed by \evtgen~\cite{Lange:2001uf}, in which final-state radiation is generated using \photos~\cite{Golonka:2005pn}. 
The interaction of the generated particles with the detector and its response are implemented using the \geant toolkit~\cite{Allison:2006ve, *Agostinelli:2002hh}, as described in Ref.~\cite{LHCb-PROC-2011-006}. 
Processing, reconstruction and selection are identical for simulated events and data. 
The simulation is used to determine correction factors for the detector acceptance and resolution as well as for quantifying background contributions and reconstruction performance.

The measurements are compared to predictions of two classes of generators, those that have not been optimized using LHC data and those that have. 
The former includes the Perugia~0 and Perugia~NOCR~\cite{Skands:2009zm} tunes of \pythia 6, both of which rely on CTEQ5L~\cite{Lai:1999wy} PDFs, and the \phojet event generator~\cite{Phojet:1995a}. 
\phojet describes soft-particle production by relying on the dual-parton model~\cite{Capella1994225}, which comprises semi-hard processes modelled by parton scattering and soft processes modelled by pomeron exchange. 
\pythia 8~\cite{Sjostrand:2007gs} is available in both classes. 
An early version of \pythia 8 is represented by version 8.145. 
In more recent versions, the default configuration has been changed to Tune~4C, which is based on \lhc measurements in the central rapidity region. 
Both \pythia 8 versions utilize the CTEQ5L PDFs. 
The results of the latest available version, \pythia 8.180, are used to represent Tune~4C. 
\pythia 8.180, together with recent versions of \herwig~\cite{Bahr:2008pv}, represent the class of recent event generators. 
In contrast to the \pythia generator, where hadronisation is described by the Lund string fragmentation, the \herwig generator relies on cluster fragmentation and the preconfinement properties of parton showers. 
Predictions of two versions of \herwig are chosen, each operated in the minimum-bias configuration, which uses the respective default underlying-event tune. 
For \herwig version 2.6.3, this corresponds to tune UE-EE-4-MRST (UE-4), while version 2.7.0~\cite{Bellm:2013lba} relies on tune UE-EE-5-MRST (UE-5). 
Both tunes were also optimized to reproduce \lhc measurements in the central rapidity region and rely on the \text{MRST LO**}~\cite{Sherstnev:2007nd} PDF set.

%% file: eventdefinition.tex

\section{Event definition and data selection}
\label{sec:Eventdefinition}

In analogy with similar approaches adopted in previous measurements~\cite{ATLAS:2010ir, Aamodt:2010pp}, an event is defined as \textit{visible} if it contains at least one charged particle in the pseudorapidity range of $2.0<\eta<4.8$ with $\pt > 0.2\gevc$ and $\ptot > 2\gevc$. 
These criteria correspond to the typical kinematic requirements for particles traversing the magnetic field and reaching the downstream tracking stations. 
In order to compare the data directly to predictions from Monte Carlo generators without having a full detector simulation, the visibility definition is based on the actual presence of real charged particles, regardless of whether they are reconstructed as tracks or not.

The tracks are corrected for detector and reconstruction effects to obtain the distribution of charged particles produced in $pp$ collisions. 
Only tracks traversing the full tracking system are considered. 
The kinematic criteria are explicitly applied to all tracks to restrict the measurement to a kinematic range in which reconstruction efficiency is high. 
The track reconstruction requires a minimum number of detector hits and a successful track fit. 
To retain high reconstruction efficiency, no additional quality requirement for suppressing the contribution from misreconstructed tracks is applied. 
To ensure that tracks originate from the primary interaction, it is required that the smallest distance of the extrapolated track to the beam line is less than $2\mm$. 
The position of the beam line is determined independently for each data taking period from events with reconstructed primary vertices. 
Additionally, a track is required to originate from the luminous region; the distance $z_{0}$ of the track to the centre of this region has to fulfil $z_{0}<3\sigma_{\text{L}}$, where the width $\sigma_{\text{L}}$ is of the order of $40\mm$, determined from a Gaussian fit to the longitudinal position of primary vertices. 
This restriction also suppresses the contamination from beam-gas background interactions to a negligible amount. 
The distribution of the $z$-position of tracks at the closest point to the beam line shows that in both high-multiplicity and single-track events, beam-gas interactions are distributed over the entire $z$-range of the VELO, whereas the distribution of tracks originating from $pp$ collisions peaks in the luminous region. 
There is no explicit requirement for a reconstructed primary vertex in this analysis. 
Together with the chosen definition of a visible event, this allows the measurement to also be performed for events with only single particles in the acceptance. 

%% file: analysis.tex

\section{Analysis}
\label{sec:AnalysisStrategy}
The measured particle multiplicity distributions and mean particle densities are corrected in four steps: 
(1) reconstructed events are corrected on an event-by-event basis by weighting each track according to a purity factor to account for the contamination from reconstruction artefacts and non-prompt particles; 
(2) the event sample is further corrected for unobserved events that fulfil the visibility criteria but in which no tracks are reconstructed; 
(3) in order to obtain measurements for single $pp$ collisions, a correction to remove pile-up events is applied; 
(4) the effects of various sources of inefficiencies, such as track reconstruction, are addressed. 
\newline
While correction factors for the multiplicity distributions and mean particle densities are the same, their implementation differs and is discussed in the following.

\subsection{Correction for reconstruction artefacts and non-prompt particles}
\label{sec:BgContamination}
The selected track sample includes three significant categories of impurities: 
approximately $6.5\,\%$ are fake tracks, less than $1\,\%$ are duplicate tracks and about $4.5\,\%$ are tracks from non-prompt particles. 
The individual contributions are determined using fully simulated events. 
Henceforth, all impurity categories are collectively referred to as background tracks. 

The probability of reconstructing a fake track, $\mathcal P_{\text{fake}}$, is dependent on the occupancy of the tracking detectors and on the track parameters. 
The occupancy dependence is determined as a function of the track multiplicity measured by the VELO and as a function of the number of hits in the downstream tracking stations. 
This accounts for the increasing probability of reconstructing a fake track depending on the number of hits in each of the tracking devices involved. 
$\mathcal P_{\text{fake}}$ also depends on $\eta$ and \pt; this is taken into account in an overall four-dimensional parametrisation. 

Duplicate tracks are reconstruction artefacts, they have only a weak dependence on tracking-detector occupancy but exhibit a pronounced kinematic dependence. 
The probability of reconstructing a duplicate track, $\mathcal P_{\text{dup}}$, is estimated as a function of $\eta$, \pt and VELO track multiplicity.

The probability that a non-prompt particle is selected, $\mathcal P_{\text{sec}}$, is also estimated as a function of the same variables as for duplicate tracks. 
The predominant contribution is due to material interaction, such as photon conversion, and depends on the amount of material traversed in the detector. 
Low $\pt$ particles are more affected. 

For each track, a combined impurity probability, $\mathcal P_{\text{bkg}}$, is calculated, which is the sum of the three contamination types, $\mathcal P_{\text{bkg}}=\mathcal P_{\text{fake}}+\mathcal P_{\text{dup}}+\mathcal P_{\text{sec}}$, and depends on the kinematic properties of the track, the occupancy of the tracking detectors and the track multiplicity. 
When measuring the mean particle densities, it is sufficient to assign a per-track weighting factor of $(1-\mathcal P_{\text{bkg}})$ to correct for the impurities mentioned above. 
However, correcting particle multiplicity distributions in the same way would lead to non-physical fractional event multiplicities. 
To obtain the background-subtracted multiplicity distributions, the procedure described below is applied. 
The description only corresponds to the full kinematic range, but the procedure is performed in each of the $\eta$ and $\pt$ sub-ranges separately. 
The impurity probability, $\mathcal P_{\text{bkg}, i}$, of each track, is summed for all tracks in an event to obtain a total event impurity correction, $\mu_{\text{ev}}$. 
This corresponds to a mean number of expected background tracks in the event and permits to calculate the probability to reconstruct a certain number of background tracks in each event, assuming Poisson statistics. 
The number of background tracks $k$ in an event with $n_{\text{ev}}$ observed tracks obeys the probability distribution 
\begin{equation}
 \mathcal P_{\text{bkg}}(\mu_{\text{ev}},k)=\frac{\mu_{\text{ev}}^{k}}{k!} e^{-\mu_{\text{ev}}}, \; \; \text{with} \; \;\mu_{\text{ev}}=\sum_{i=1}^{n_{\text{ev}}} \mathcal P_{\text{bkg}, i}. 
\label{eq:poisson}
\end{equation}
From this relation we derive the probability that an event contains a given number of real prompt particles. 
Summing the normalized probability distribution of all events we obtain the multiplicity distribution corrected for background tracks.

\subsection{Correction for undetected events}
\label{sec:NotRecEvent}
Defining a visible event based on the properties of the actual charged particles present in the event rather than on the reconstructed tracks introduces a fraction of spuriously undetected events. 
These are events that should be visible but contain no reconstructed tracks and thus remain undetected. 
These \textit{unobserved} events are most likely to occur when few charged particles are within the kinematic acceptance. 
The reconstruction of a track can fail due to multiple scattering, material interaction, or inefficiencies of the detector or of the reconstruction algorithms. 
In order to determine the amount of undetected events that nevertheless fulfil the visibility definition, a data-driven approach is adopted.

The true multiplicity distribution for visible events, $T(n)$, where $n$ is the number of charged particles, starts at $n=1$. 
Since some of these events have no reconstructed tracks, they follow a multiplicity distribution $U(n)$ starting from $n=0$. 
As an event can only be detected if at least one track is reconstructed, $U(0)$ cannot be determined directly. 
However, the number of undetected events can be estimated from the observed uncorrected distribution $U(n)$, if the average survival probability, $\mathcal{P}_{sur}$, for a single particle in the kinematic acceptance is known. 
Assuming that the survival probability, which is determined from simulation, is independent for two or more particles, the observed distribution is approximated in terms of the still unknown actual multiplicity distribution $T$
\begin{equation}
 U(k) = \sum\limits_{n\geq k} \binom{n}{k} \mathcal{P}_{sur}^{k} (1-\mathcal{P}_{sur})^{n-k} T(n).
\label{eq:AllBins}
\end{equation}
This equation is only valid under the assumption that reconstruction artefacts, such as fake tracks, which increase the number of observed tracks with respect to the number of true tracks, can be ignored.
Following this approach, an event with a certain number of particles is only reconstructed with the same number of tracks or fewer, but not with more tracks. 
The uncertainties due to these assumptions are evaluated in simulation and are accounted for as systematic uncertainties. 
Equation~\ref{eq:AllBins} allows $U(0)$ to be estimated from the true distribution $T$. 
All actual elements $T(k)$ can also be expressed using the corresponding uncorrected measured bin $U(k)$ and correction terms of $T(n)$ at higher values of $n > k$,
\begin{equation}
\begin{split}
 & U(0) \approx \displaystyle\sum\limits_{k=1}^{\rm{r}} (1-\mathcal{P}_{sur})^{k} T(k) \text{ \; with} \\
 & T(k) \approx \frac{U(k)}{\mathcal{P}_{sur}^{k}} - \displaystyle\sum\limits_{n=k+1}^{k+\rm{r}} \binom{n}{k} (1-\mathcal{P}_{sur})^{n-k} T(n).\\
\end{split}
\label{eq:Msolution}
\end{equation}
Combining the formulas in Eq.~\ref{eq:Msolution} results in a recursive expression for $U(0)$, which can be calculated numerically up to a given order $r$. 
The procedure is tested in simulation, where the estimated and actual fractions of undetected events agree within an uncertainty of $13\,\%$. 
This is considered as a systematic uncertainty related to the assumptions made in the calculation. 
The fraction of undetected events obtained for data is $2.3\,\%$ compared to $3.1\,\%$ in simulation. 
The fraction estimated in data is added to the measured multiplicity distributions and is also considered in the event normalisation of the mean particle density measurement.

\subsection{Pile-up correction}
\label{sec:PileupCorrection}
The average number of interactions per bunch crossing in the selected data taking period is small, resulting in a limited bias from pile-up.
The measured particle multiplicity distributions are mainly composed of single $pp$ collisions and a small fraction of additional second $pp$ collisions.  
Therefore events with larger pile-up can be neglected. 
To obtain the particle multiplicity distribution of single $pp$ collisions the iterative approach used in Ref.~\cite{LHCb-PAPER-2011-011} is applied. 
The procedure typically converges after two iterations when the change of the multiplicity distribution is of the order of the statistical uncertainty. 
The pile-up correction changes the mean value of the multiplicity distribution by $3.3\,\%$. 
The measurements of the mean particle density are normalised to the total number of $pp$ collisions. 

\subsection{Efficiency correction and unfolding procedure}
\label{sec:EfficiencyUnfolding}
The final correction step accounts for limited efficiencies due to detector acceptance $(\epsilon_{\text{acc}})$ in the kinematic range of $2.0<\eta<4.8$ and 
track reconstruction $(\epsilon_{\text{tr}})$. 
For particles fulfilling the kinematic requirements, the detector acceptance describes the fraction that reach the end of the downstream tracking stations and are unlikely to interact with material or to be deflected out of the detector by the magnetic field. 
This fraction and the overall reconstruction efficiency are evaluated independently using simulated events. 
Correction factors are determined as functions of pseudorapidity and transverse momentum. 
No multiplicity dependence is observed. 
The mean particle densities are corrected by applying a combined correction factor of $1/(\epsilon_{\text{acc}} \epsilon_{\text{tr}})$ to each track in the same way as described in Sect.~\ref{sec:BgContamination}. 

In order to correct the particle multiplicity distributions, an unfolding technique based on a detector response matrix is employed. 
The response matrix, $R_{m,n}$, accounts for inefficiencies due to the detector acceptance and track reconstruction. 
It is constructed from the relation between the distribution of true prompt charged-particles $T(n)$ and the distribution of measured tracks $M(m)$, subtracted for background and pile-up, 
\begin{equation}
 M(m) = \sum_{n} R_{m,n} T(n).
\end{equation}
The matrix is obtained from simulated events. 
The simulated number of charged particles per event, $n$, is compared to the corresponding number of reconstructed and background subtracted tracks, $m$. 
Thus each possible value of simulated particle multiplicity is mapped to a distribution of reconstructed tracks. 
For very high multiplicities, the available number of events from the Monte Carlo sample is not sufficient to populate the entire matrix. 
The mapping is well described by a Gaussian distribution with mean value $\bar{m}$ and standard deviation $\sigma_{m}$. 
The distribution of $\bar{m}$ and $\sigma_{m}$ for a true multiplicity bin $n$ can be parametrized by combinations of polynomial and logarithmic functions. 
This allows an extrapolation of the matrix up to large values of $n$ and simultaneously suppresses the effect of statistical fluctuations in the entries of the matrix. 
For further information the reader is referred to the Appendix, where an example of the detector response matrix is shown in Fig.~\ref{fig:ResponseMatrix}. 

To extract the true particle multiplicity distribution $T(n)$ from the measured distribution $M(m)$, a procedure based on $\chi^{2}$-minimization~\cite{Blobel:157405, Zech:1995he} of the measured distribution $M(m)$ and the folded distribution $R_{m,n}\tilde{T}(n)$ for different hypotheses of the true distribution, $\tilde{T}(n)$, is adopted. 
The range of variation of $\tilde{T}(n)$ is constrained by parametrising the multiplicity distributions. 
To avoid introducing model dependencies to the unfolded result, six different models with up to eight floating parameters are used. 
Five models are based on sums of exponential functions combined with polynomial functions of various order in the exponent and as a multiplier. 
In addition, a model based on a sum of negative binomial distributions is used. 
While particle multiplicities in $\eta$ and $\pt$ bins can be well described by two negative binomial distributions, this is not sufficient for the multiplicity distribution in the full kinematic range, where this model has not been employed. 
All the parametrisations used are capable of describing the simulated multiplicity distributions. 
The floating parameters of the hypothesis $\tilde{T}(n)$ are varied in order to minimise the $\chi^{2}$-function
\begin{equation}
 \chi^2(\tilde{T}) = \displaystyle\sum\limits_{m} \frac{1}{E(m)^{2}}\left(M(m)-\displaystyle\sum\limits_{n} R_{mn}\tilde{T}(n)\right)^2,
\label{eq:chi2}
\end{equation}
where $E(m)$ represents the uncertainty of the measured distribution $M(m)$. 
The parametrisation model yielding the best $\chi^{2}$-value is chosen as the central result, the other models are considered in the systematic uncertainty determination. 
Both the binned and total event unfolding procedures using simulated data are found to reproduce the generated distributions satisfactorily. 
The uncertainty of the unfolded distribution is determined through pseudo-experiments. 
Each pseudo-experiment is generated from the analytical model with the parameters randomly perturbed according to the best fit and the correlation matrix.

As a consistency check, a Bayesian unfolding technique~\cite{D'Agostini:1994zf} is used. 
The unfolded distributions of both methods in all kinematic bins are found to be in agreement.

The unfolded distribution for the total event is truncated at a value of 50 particles and the binned distributions at a value of 20 particles. 
This corresponds to the limit where, even with the extended detector-response matrix, larger particle multiplicities cannot be fully mapped to the range of the measured track-multiplicity distribution and where systematic uncertainties become large. 

%% file: systematics.tex

\section{Systematic uncertainties}
\label{sec:Systematics}
The precision of the measurements of charged particle multiplicities and mean particle densities are limited by systematic effects. 
The bin contents of the particle multiplicity distribution for the full event typically have a relative statistical uncertainty in the range of $10^{-4}$ to $10^{-2}$ for low and high multiplicities, respectively. 
The systematic uncertainties are typically around $1-10\,\%$, the largest contribution arising from the uncertainty of the amount of detector material. 
All individual contributions are discussed below.

The properties of fake tracks are studied in detail by using fully simulated events. 
The agreement between data and simulation is verified by estimating the fake-track fraction in both samples by probing the matching probability of track segments in the long-track reconstruction algorithm. 
The results are in good agreement and the differences amount to an overall $2\,\%$ systematic uncertainty on the applied correction factors.

The systematic uncertainty introduced by differences in the fraction of duplicate tracks in data and simulation is determined by studying the number of track pairs with small opening angles. 
The observed excess of duplicate tracks in data results in a relative systematic uncertainty on the duplicate-track fraction of $9\,\%$. 
As the total amount of this type of reconstruction artefacts is small, this results in an overall $0.1\,\%$ systematic uncertainty on the final result.

Uncertainties introduced by the correction for non-prompt particles depend predominantly on the knowledge of the amount of material within the detector. 
The agreement with the amount of material modelled in the simulation, on average, is found to be within $10\,\%$. 
In order to estimate the effects of non-prompt particles still passing the track selection, their composition is studied. 
Around $40\,\%$ of the wrongly selected particles arise from photon conversion and is related to the uncertainty of the amount of material. 
Another third of the particles are decay products of $K_{\text{S}}^{\text{0}}$ mesons, whose production cross-section has previously been measured by \lhcb~\cite{LHCb-PAPER-2010-001} to be in good agreement with simulation. 
Around $20\,\%$ of the particles originate from decays of $\Lambda$ baryons and hyperons. 
These are measured to disagree by approximately $40\,\%$ with the production cross-sections used in the simulation. 
Combining these contributions results in a $12\,\%$ systematic uncertainty on the fraction of non-prompt particles.

To account for differences between the actual track reconstruction efficiency and that estimated from simulation, a global systematic uncertainty of $4\,\%$ in average is assigned~\cite{Jaeger:1402577, LHCb-PAPER-2010-002}. 

The uncertainty on the detector acceptance can be split in two components: 
the uncertainty on the knowledge of the detector material and the uncertainty related to the requirement for particles to have trajectories within the acceptance of the downstream tracking stations. 
The momentum distributions of charged particles in data and in simulation are in good agreement, therefore the second effect is negligible. 
The remaining uncertainty related to material interaction leads to a relative systematic uncertainty on the correction factors of $3\,\%$ and is assigned as an individual factor for each track. 

A modified response matrix is used to estimate the impact on the multiplicity distributions of systematic uncertainties due to the track reconstruction and detector acceptance. 
The systematic uncertainties of both efficiencies are combined quadratically and result in a $5\,\%$ uncertainty on the response matrix.
A response matrix with an efficiency decreased by this value is generated. 
The whole unfolding procedure (Sect.~\ref{sec:EfficiencyUnfolding}) is repeated with this matrix and the full difference to the nominal result is assigned as uncertainty.

Model dependencies due to the parametrisations used to unfold the true particle multiplicity distributions are determined by sampling six different parametrisation models for each of the multiplicity distributions. 
The model corresponding to the minimum $\chi^{2}$ value of the unfolding fit is taken as the central result, while the maximum difference in each bin between all models and the central result is taken as the systematic uncertainty. 
This difference is small compared to the uncertainty due to the modified response matrix.

Uncertainties related to the correction for undetected events (Sect.~\ref{sec:NotRecEvent}) are dominated by the $13\,\%$ systematic uncertainty arising from the assumptions made in the calculation model. 
In addition, the average survival probability used in this model is affected by uncertainties of the amount of detector material, detector acceptance and track reconstruction efficiency. 
This sums to a maximum uncertainty of $15\,\%$ on the number of undetected events. 
Only bins from one to three tracks are affected, where the variation is dominated by this uncertainty. 
For the particle densities, the impact is negligible with respect to other uncertainties. 
For the particle multiplicity distributions it results in a small change of $0.4\,\%$ of the truncated mean.

Uncertainties related to the pile-up fraction are evaluated to be negligible compared to all other contributions as the total size of the corrections is already small.

The effect of non-zero beam crossing angles is determined to be insignificant, as well as the background induced by beam gas interactions.

%% file: results.tex

\section{Charged particle densities}
\label{sec:ResultsDensity}

\begin{figure}[t]
\begin{center}
  \subfigure{\includegraphics*[width=0.49\textwidth]{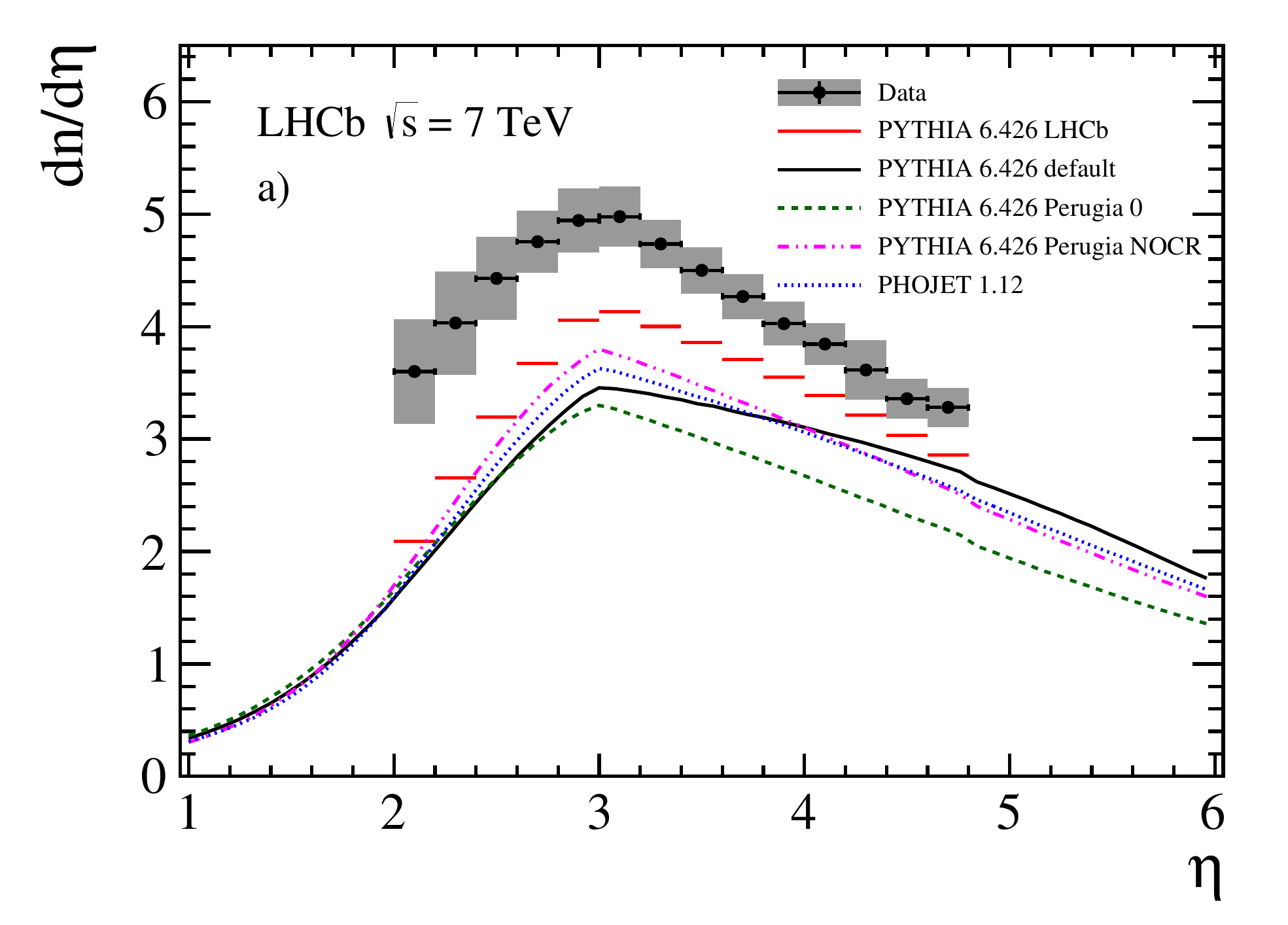}}
  \subfigure{\includegraphics*[width=0.49\textwidth]{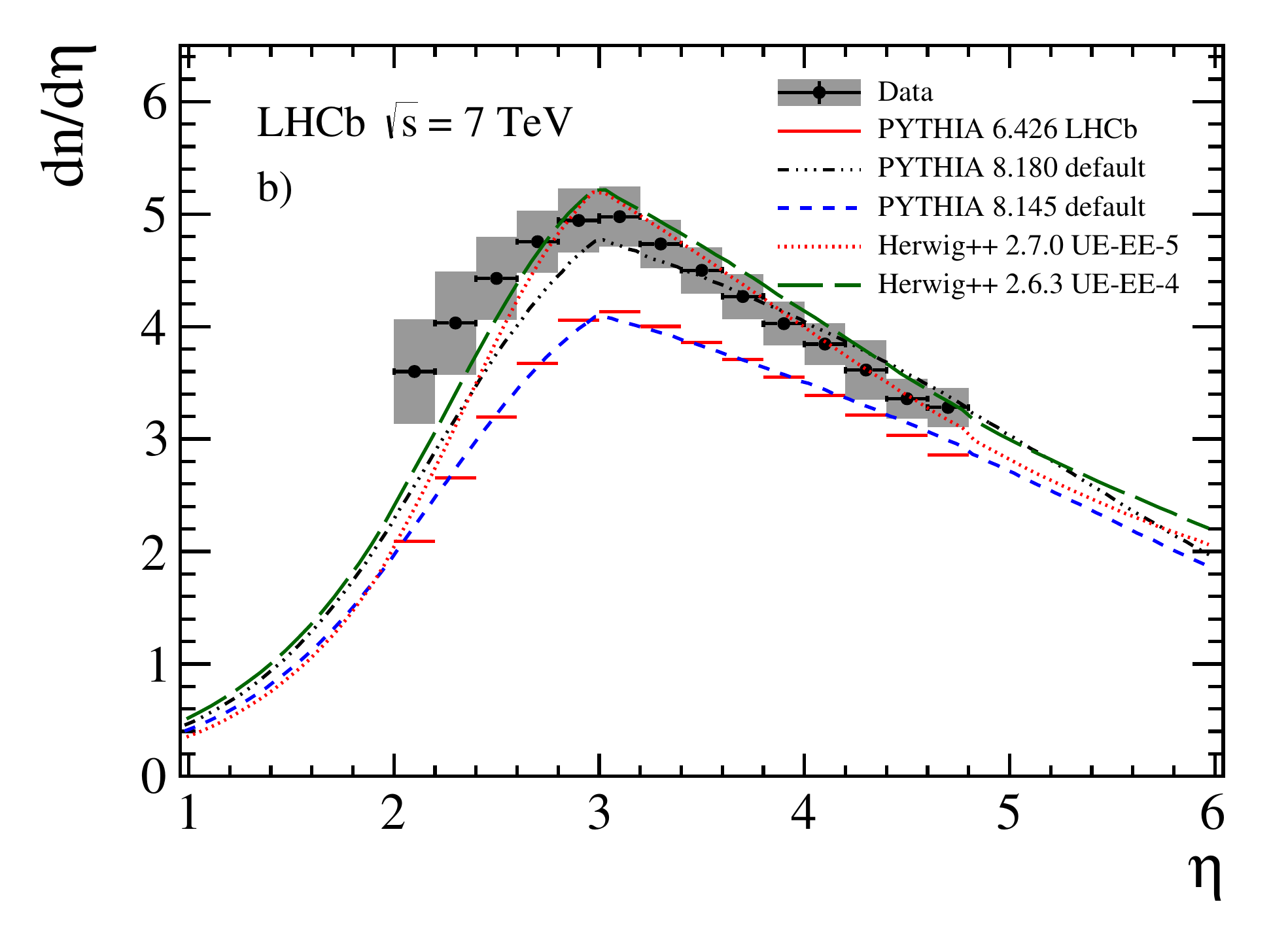}}
\end{center}
\caption{\small Charged particle density as a function of $\eta$. The \lhcb data are shown as points with statistical error bars (smaller than the marker size) and combined systematic and statistical uncertainties as the grey band. The measurement is compared to several Monte Carlo generator predictions, (a) \pythia 6 and \phojet, (b) \pythia 8 and \herwig. Both plots show predictions of the \lhcb tune of \pythia 6, which is used in the analysis.}
\label{fig:DensityEta}%
\end{figure}

The fully corrected measurement of mean particle densities in the kinematic region of $p>2\gevc$, $\pt>0.2\gevc$ and $2.0<\eta<4.8$ is presented as a function of pseudorapidity in Fig.~\ref{fig:DensityEta} and as a function of transverse momentum in Fig.~\ref{fig:DensityPt}; the corresponding numbers are presented in the Appendix. 
The data points show a characteristic drop towards larger pseudorapidities but also a falling edge for $\eta<3$, which is caused by the minimum momentum requirement in this analysis. 
This is qualitatively described by all considered Monte Carlo event generators and their tunes.

The first group of generators that are compared to our measurements are different tunes of \pythia 6 and \phojet and are shown in Figs.~\ref{fig:DensityEta}a and~\ref{fig:DensityPt}a. 
The default configuration of \pythia 6.426 underestimates the amount of charged particles from roughly $20\,\%$ at large $\eta$ up to $50\,\%$ at small $\eta$. 
The descending slopes towards small and large pseudorapidities are also insufficiently modelled. 
The Perugia~NOCR tune shows a slight improvement in shape and in the amount of charged particles; Perugia~0 predicts an even smaller mean particle density over the whole kinematic range. 
Predictions of the \phojet generator are similar to the tunes of \pythia 6. 
In this group of predictions, the LHCb tune of \pythia 6 provides the best agreement with the data but still underestimates the charged-particle production rate by $10-40\,\%$. 
This behaviour is also observed in the \pt dependence, where all configurations underestimate the number of charged particles. 
The aforementioned generator predictions were optimized without input of LHC measurements.

\begin{figure}[t]
\begin{center}
  \subfigure{\includegraphics*[width=0.49\textwidth]{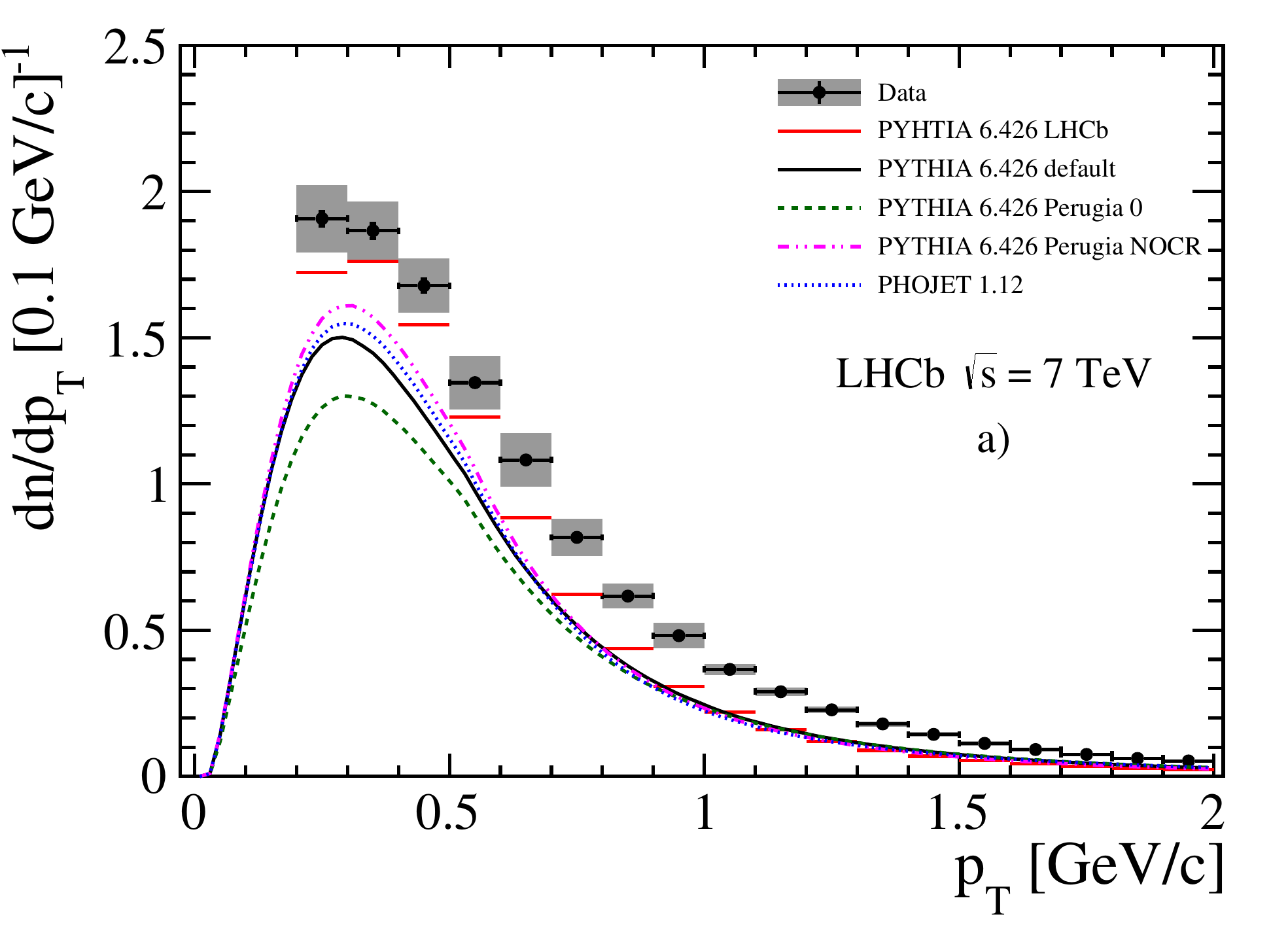}}
  \subfigure{\includegraphics*[width=0.49\textwidth]{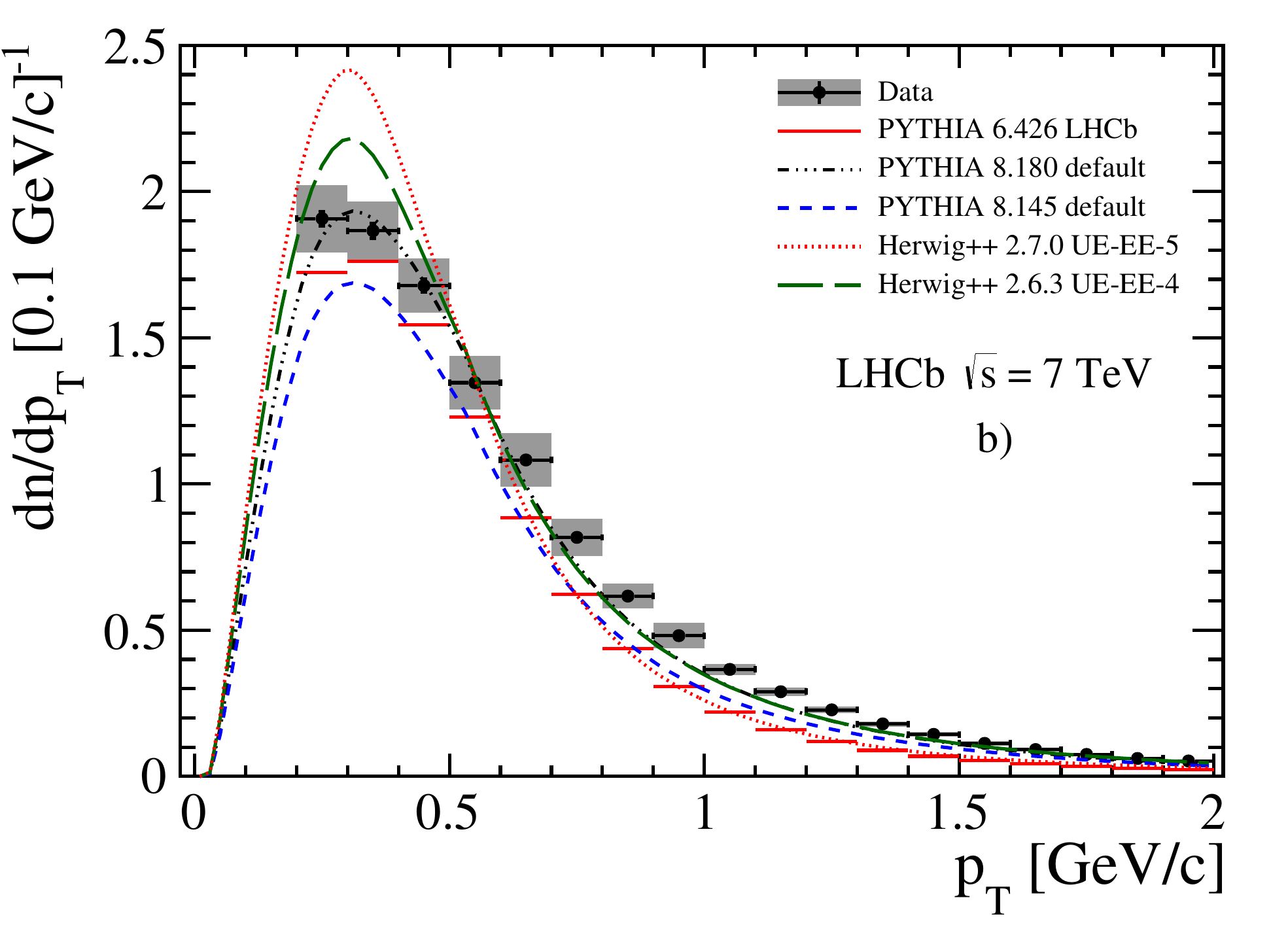}}
\end{center}
\caption{\small Charged particle density as a function of \pt. The \lhcb data are shown as points with statistical error bars (smaller than the marker size) and combined systematic and statistical uncertainties as the grey band. The measurement is compared to several Monte Carlo generator predictions, (a) \pythia 6 and \phojet, (b) \pythia 8 and \herwig. Both plots show predictions of the \lhcb tune of \pythia 6, which is used in the analysis.}
\label{fig:DensityPt}%
\end{figure}

Predictions from the more recent generators \pythia 8 and \herwig are shown in Figs.~\ref{fig:DensityEta}b and~\ref{fig:DensityPt}b. 
\pythia 8.145 with default parameters was released without tuning to \lhc measurements and is not better than the \lhcb tune of \pythia 6. 
In contrast, \pythia 8.180, which was optimized on \lhc data, describes the measurements significantly better than the previous version. 
The predictions of \herwig are also in reasonably good agreement with data, although the charged-particle production rate is underestimated at small pseudorapidities. 
The \herwig generator version 2.7.0, which uses tune UE-5, overestimates the number of prompt charged particles in the low \pt range but underestimates it at larger transverse momenta. 
The predictions of \herwig in version 2.6.3, which relies on tune UE-4, show a more complete description of the data. 
Both event generators, \pythia 8 and \herwig, describe the data over a wide range.

\begin{figure}[t]
  \begin{center}  
    \subfigure{\includegraphics*[width=0.6\textwidth]{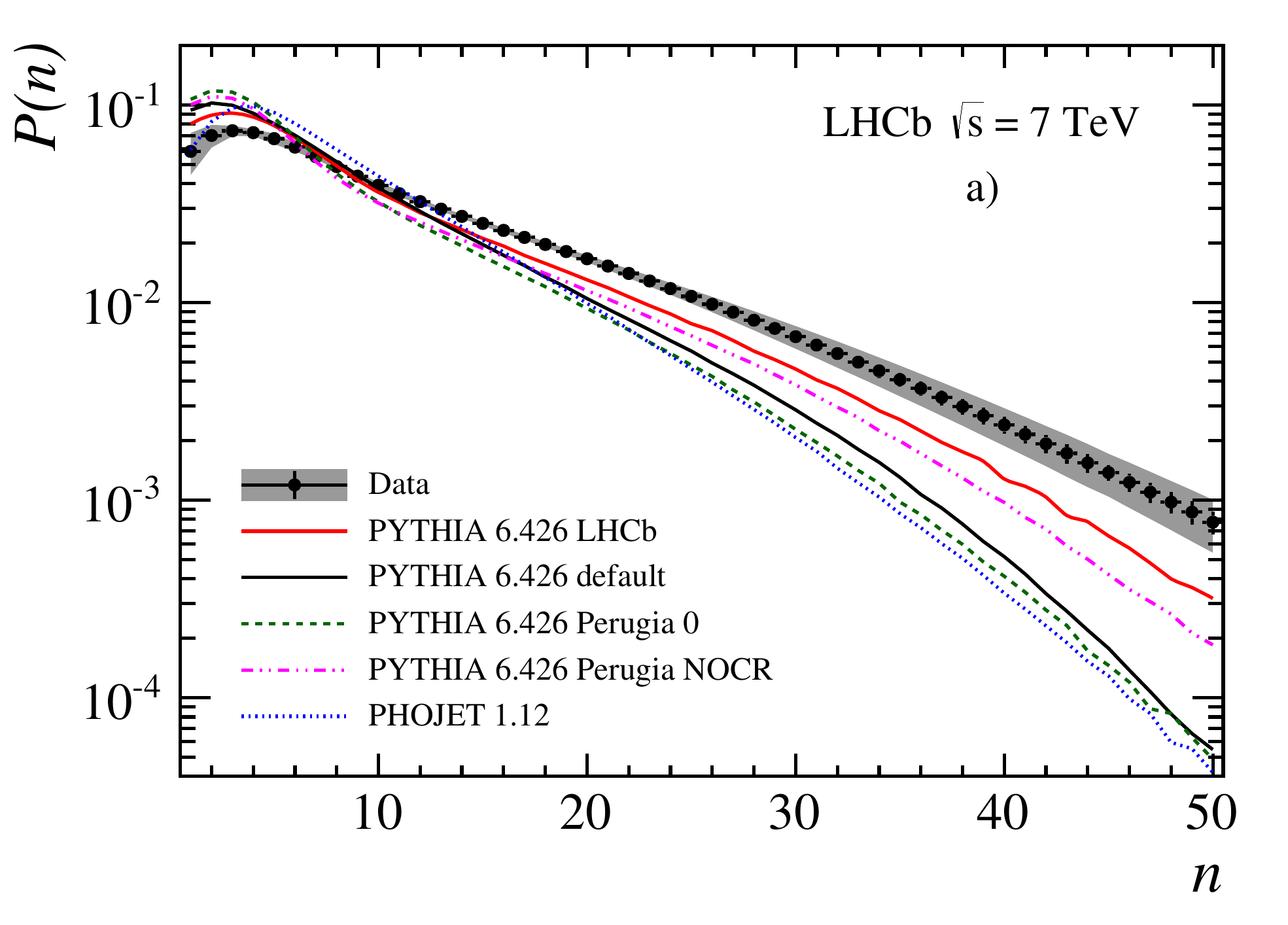}}
    \subfigure{\includegraphics*[width=0.6\textwidth]{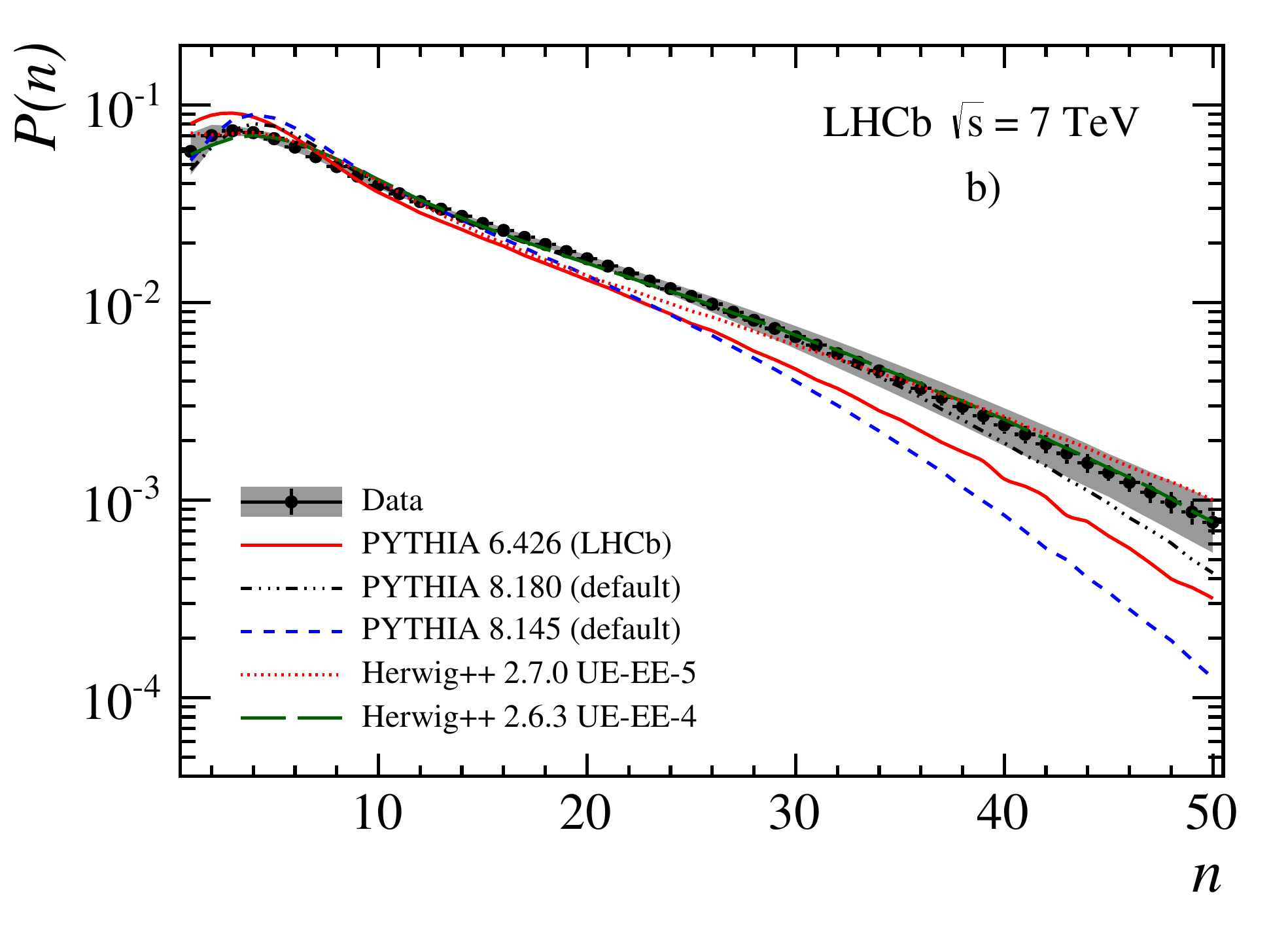}}
  \end{center}
\caption{\small Observed charged particle multiplicity distribution in the full kinematic range of the analysis. The error bars represent the statistical uncertainty, the error band shows the combined statistical and systematic uncertainties. The data are compared to several Monte Carlo predictions, (a) \pythia 6 and \phojet, (b) \pythia 8 and \herwig. Both plots show predictions of the \lhcb tune of \pythia 6, which is used in the analysis.}
\label{fig:TotalMult}%
\end{figure}

\section{Multiplicity distributions}
\label{sec:ResultsMultiplicity}
The charged particle multiplicity distribution in the full kinematic range of the analysis is shown in Fig.~\ref{fig:TotalMult}, compared to the predictions from the event generators. 
The corresponding mean value, $\mu$, and the root-mean-square deviation, $\sigma$, of the distribution, truncated in the range from 1 to 50 particles, is measured to be $\mu=11.304\pm0.008\pm0.091$ and $\sigma=9.496\pm0.006\pm0.021$, where the uncertainties are statistical and systematic, respectively. 
Using the full range gives consistent results with the value obtained from the particle densities. 
All generators that do not use \lhc data input underestimate the multiplicity distributions. 
In this comparison, the \phojet generator predicts the smallest probabilities to observe a large multiplicity event, being in disagreement with the measurement. 
This can be understood since \phojet mostly contains soft scattering events. 
All \pythia 6 tunes underestimate the charged particle production cross-section significantly. 
The prediction from the \lhcb tune is closest to the data, but the mean value of the distribution is still about $15\,\%$ too small. 
Calculations from more recent generators are in better agreement with the measurement. 
While \pythia 8.145 gives the same insufficient description of the data as its predecessor, the prediction of version 8.180 using Tune~4C shows a reasonable agreement. 
The \herwig event generator using the underlying event tune UE-4 shows good agreement with the measurement and reproduces the data better than the more recent UE-5 tune.

\begin{figure}[!h]
  \begin{center}
    \subfigure{\includegraphics*[width=0.48\textwidth]{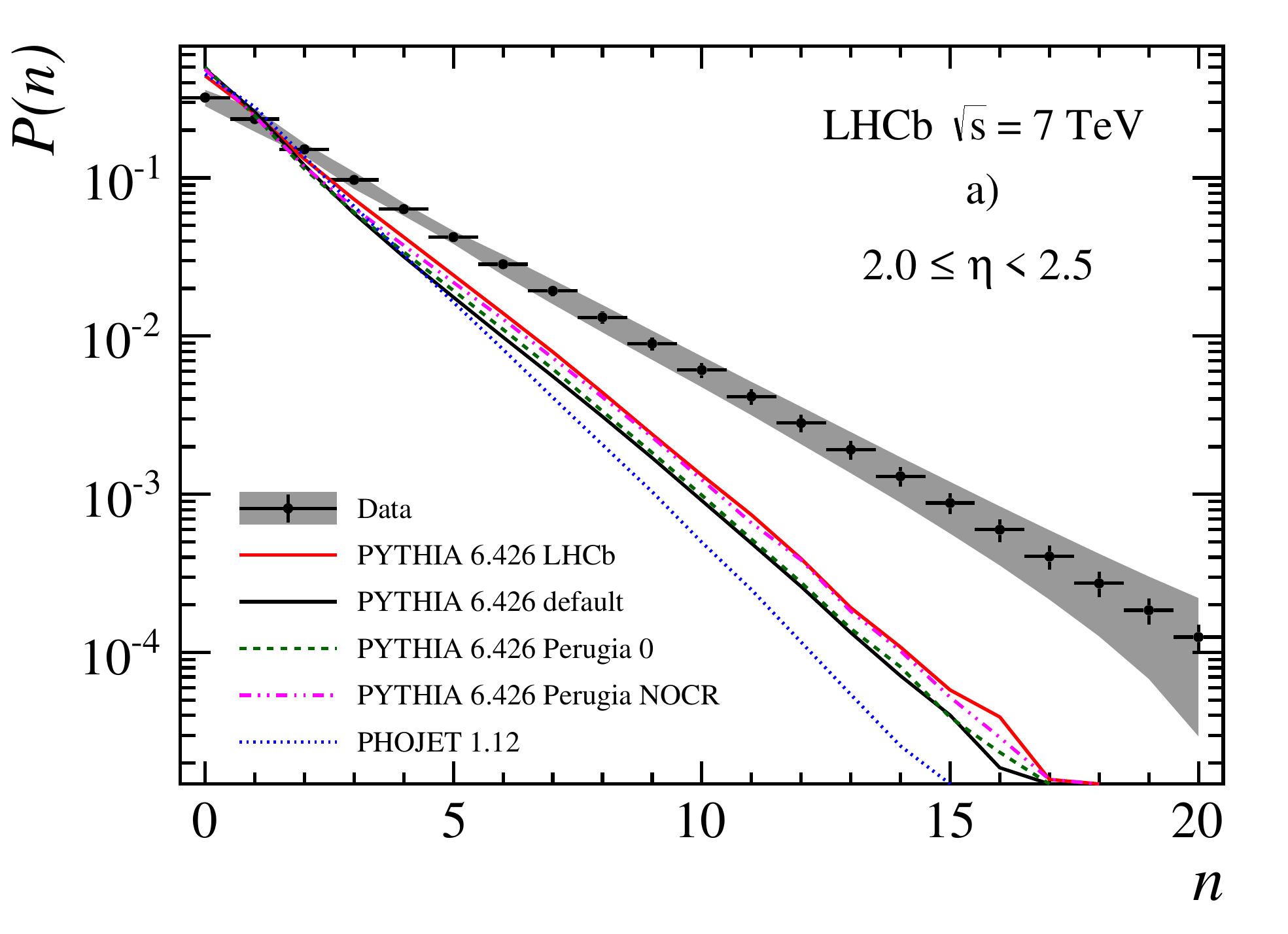}}
    \subfigure{\includegraphics*[width=0.48\textwidth]{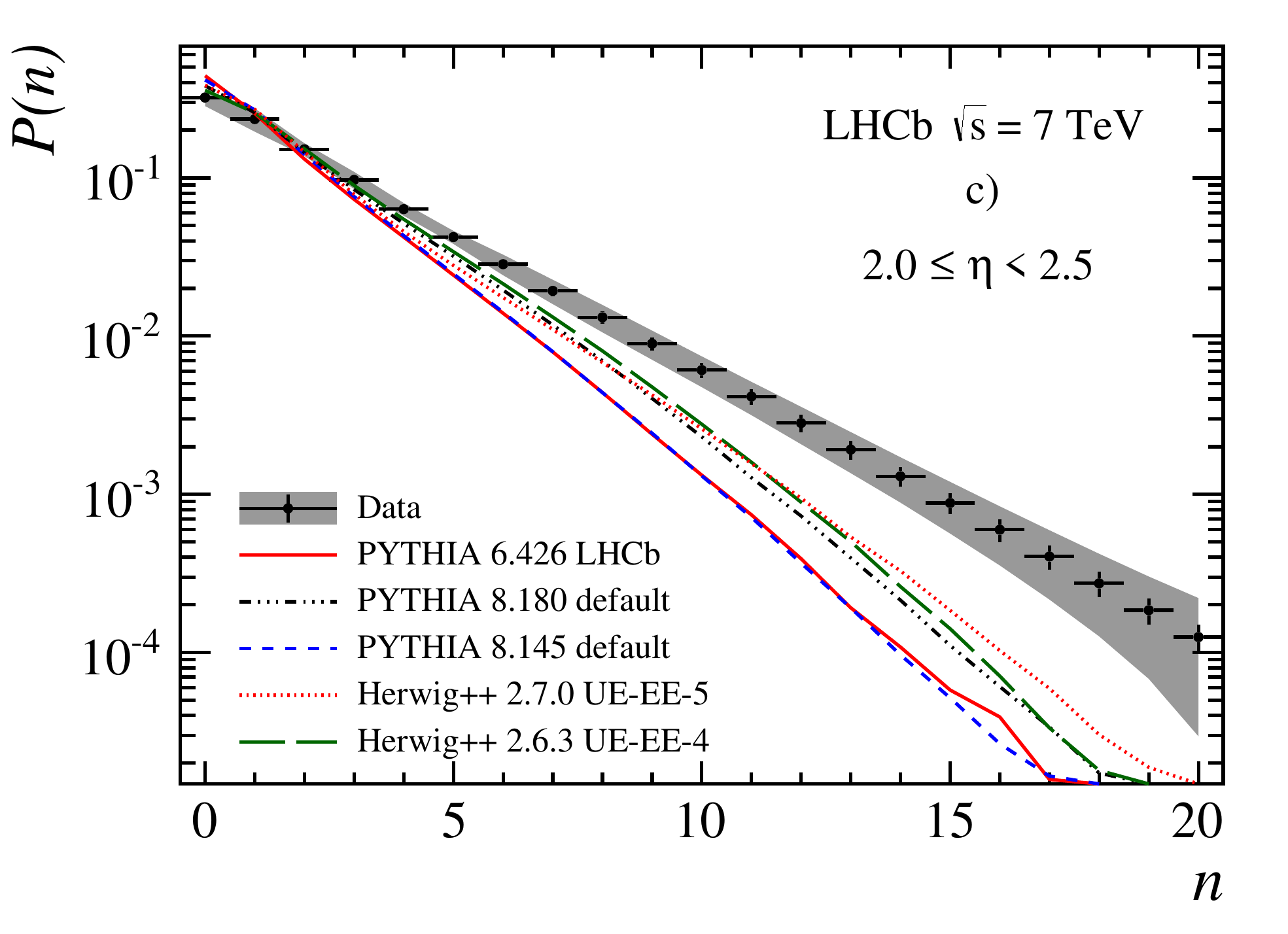}}
    \subfigure{\includegraphics*[width=0.48\textwidth]{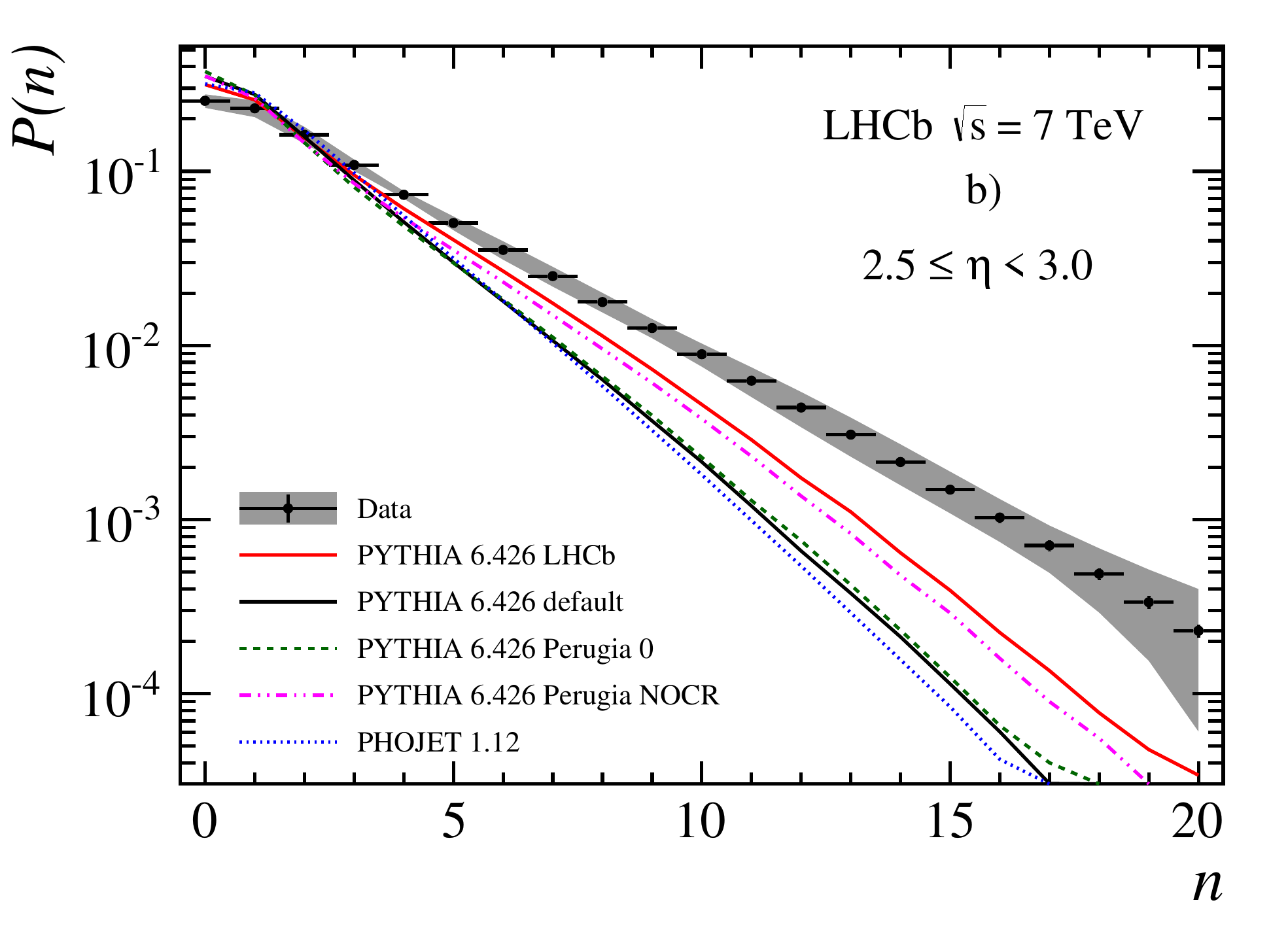}}
    \subfigure{\includegraphics*[width=0.48\textwidth]{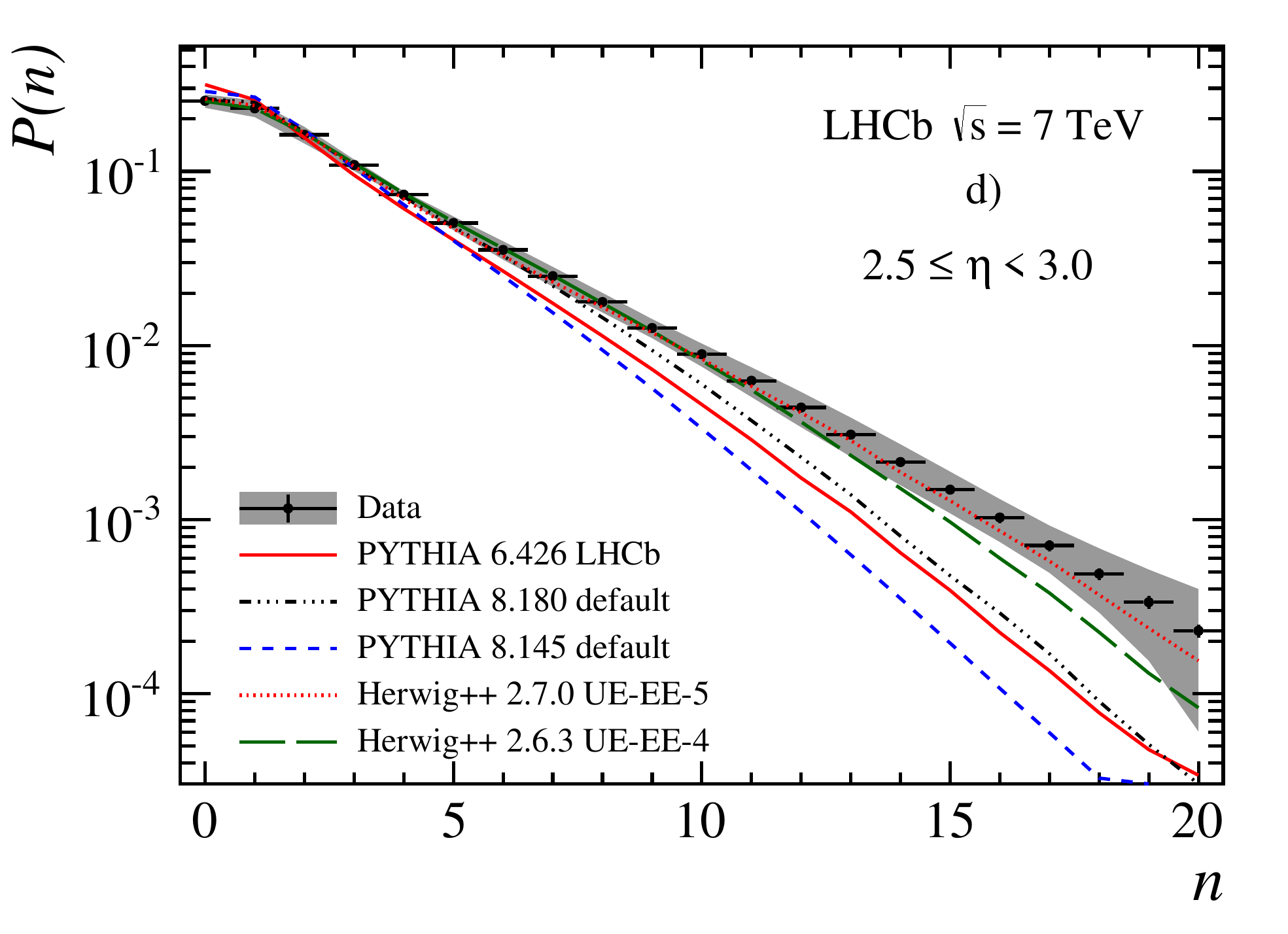}}
  \end{center}
\caption{\small Observed charged particle multiplicity distribution in different $\eta$ bins. 
Error bars represent the statistical uncertainty, the error bands show the combined statistical and systematic uncertainties. 
The data are compared to Monte Carlo predictions, (a,b) \pythia 6 and \phojet, (c,d) \pythia 8 and \herwig. 
All plots show predictions of the \lhcb tune of \pythia 6, which is used in the analysis.}
\label{fig:BinnedMult_Eta1}%
\end{figure}
\begin{figure}[h]
  \begin{center}
    \subfigure{\includegraphics*[width=0.48\textwidth]{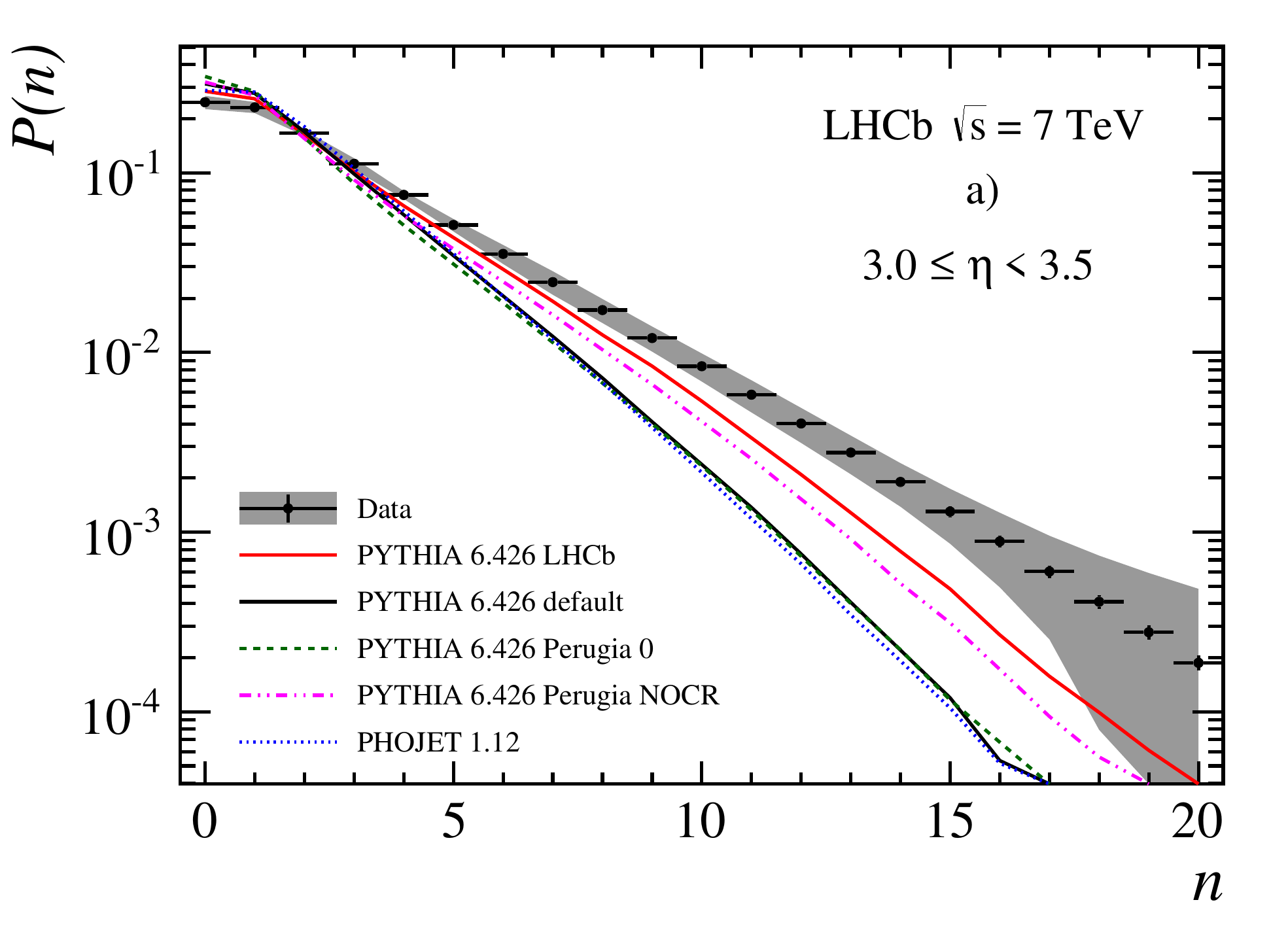}}
    \subfigure{\includegraphics*[width=0.48\textwidth]{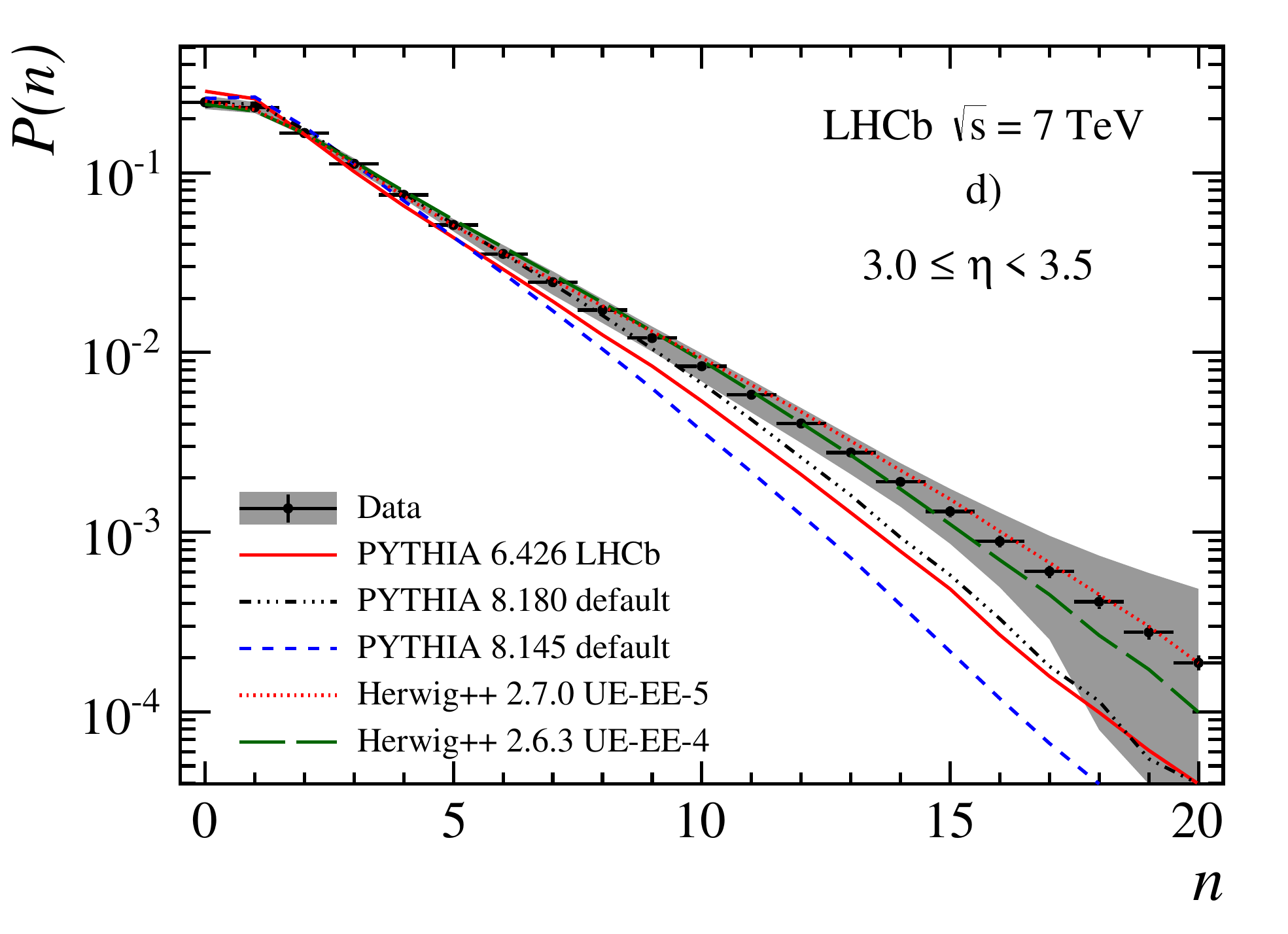}}
    \subfigure{\includegraphics*[width=0.48\textwidth]{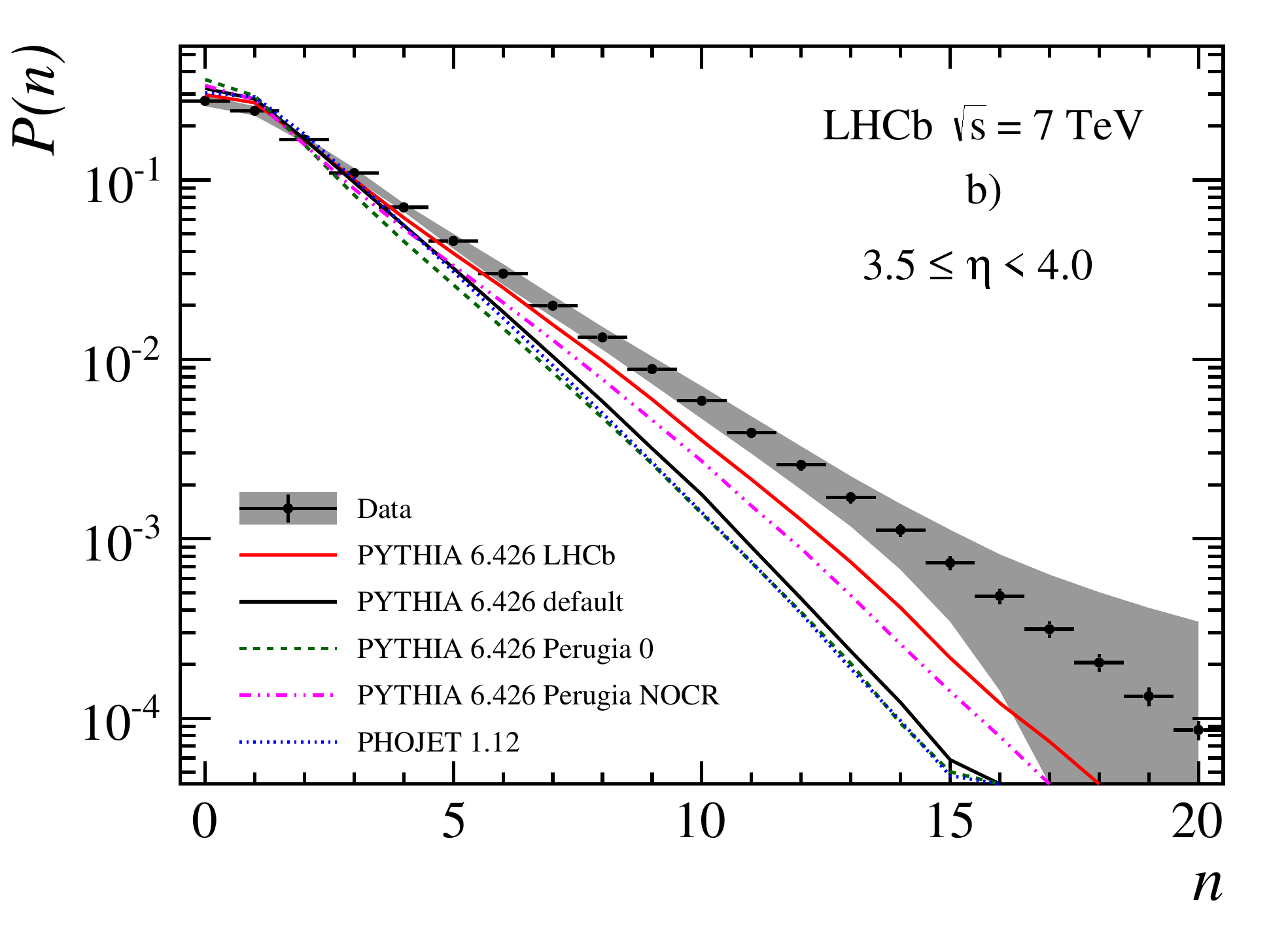}}
    \subfigure{\includegraphics*[width=0.48\textwidth]{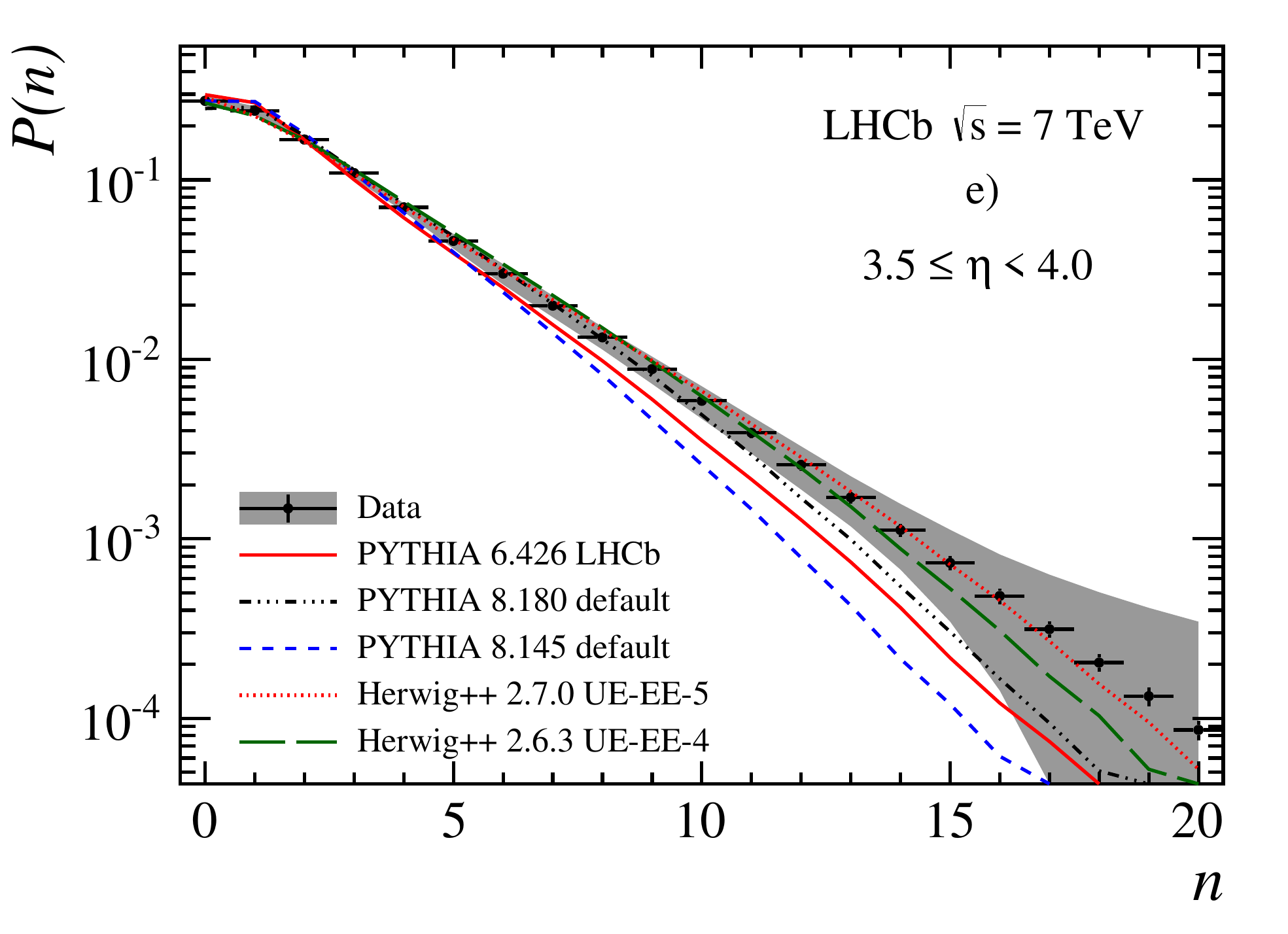}}
    \subfigure{\includegraphics*[width=0.48\textwidth]{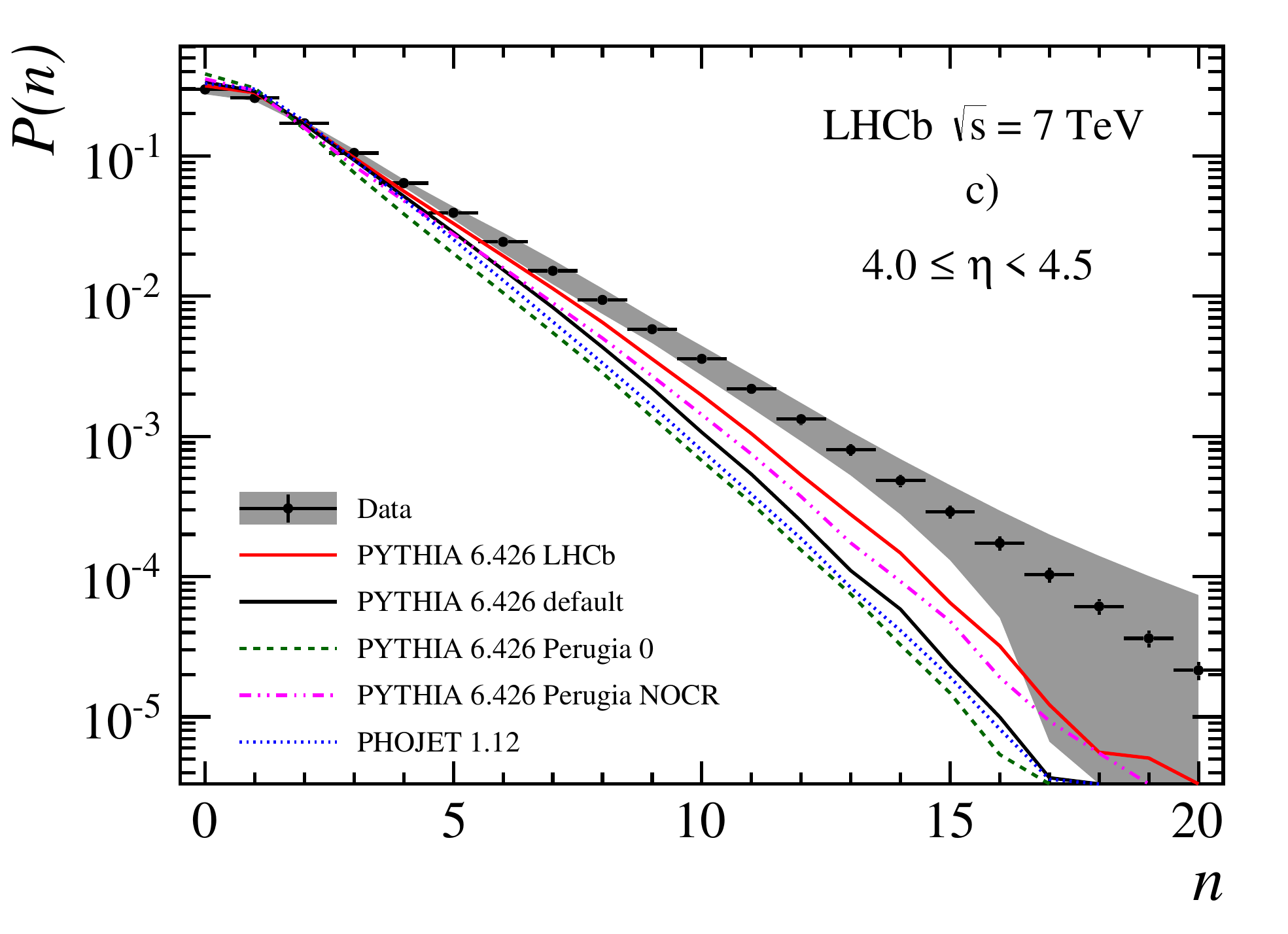}}
    \subfigure{\includegraphics*[width=0.48\textwidth]{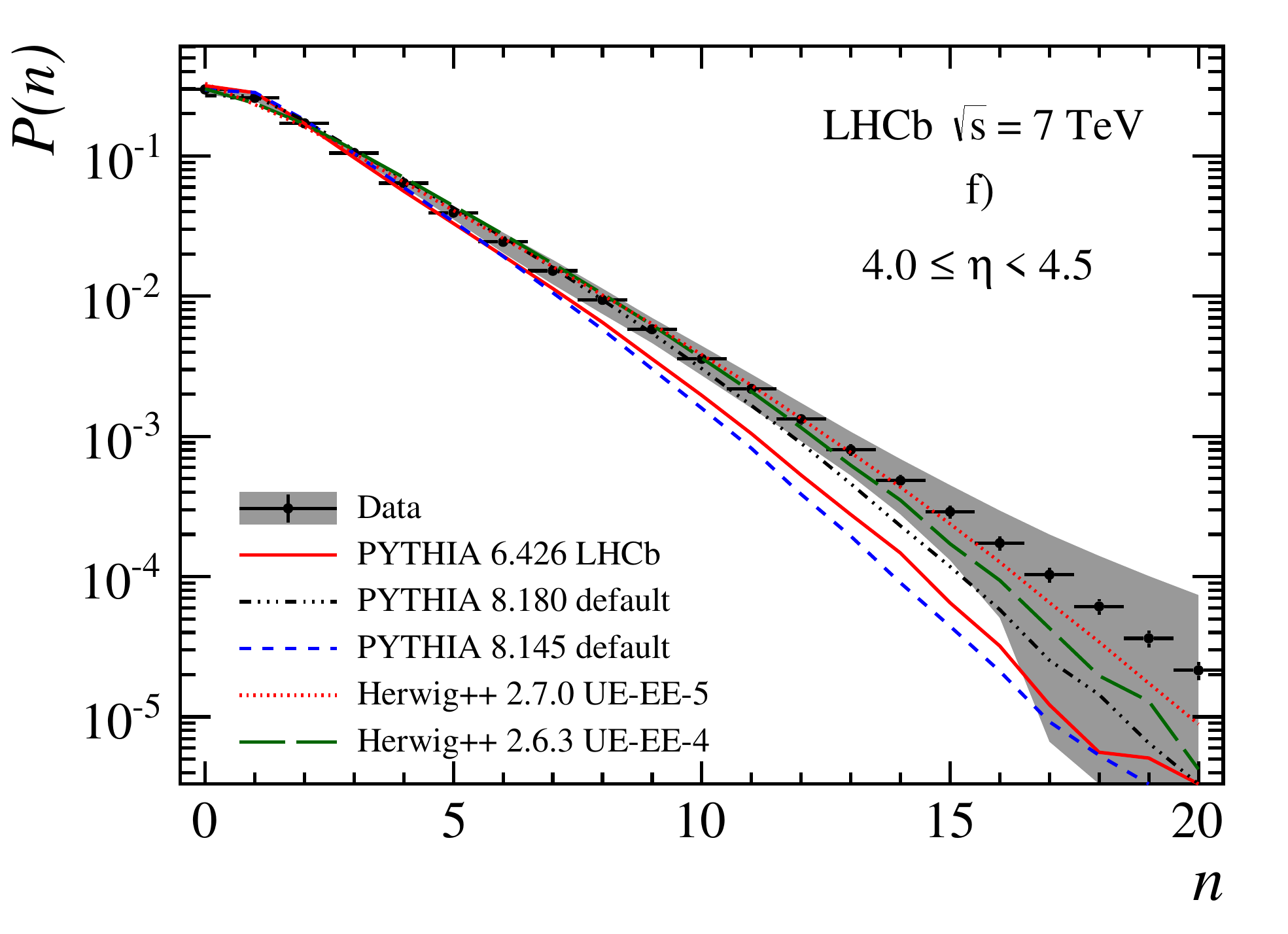}}
    \end{center}
\caption{\small Observed charged particle multiplicity distribution in different $\eta$ bins. 
Error bars represent the statistical uncertainty, the error bands show the combined statistical and systematic uncertainties.
The data are compared to Monte Carlo predictions, (a-c) \pythia 6 and \phojet, (d-f) \pythia 8 and \herwig. 
All plots show predictions of the \lhcb tune of \pythia 6, which is used in the analysis.}
\label{fig:BinnedMult_Eta2}%
\end{figure}

Charged particle multiplicity distributions for bins in pseudorapidity are displayed in Figs.~\ref{fig:BinnedMult_Eta1} and \ref{fig:BinnedMult_Eta2}. 
The comparison with the predictions from Monte Carlo generators shows the same general features as discussed for the integrated distribution. 
The predictions of \phojet and \pythia 6 all underestimate the particle multiplicity. 
The difference in particle production is most prominent at small $\eta$, where the minimum $\ptot$ requirement in this analysis significantly reduces the amount of particles. 
Even though the LHCb tune is in better agreement with the data, the difference remains large. 
Recent generator predictions match the data better. 
Both \pythia 8 and \herwig show good agreement with data at larger pseudorapidity, only the range from $2<\eta<3$ being still underestimated.

\begin{figure}[h]
  \begin{center}
    \subfigure{\includegraphics*[width=0.48\textwidth]{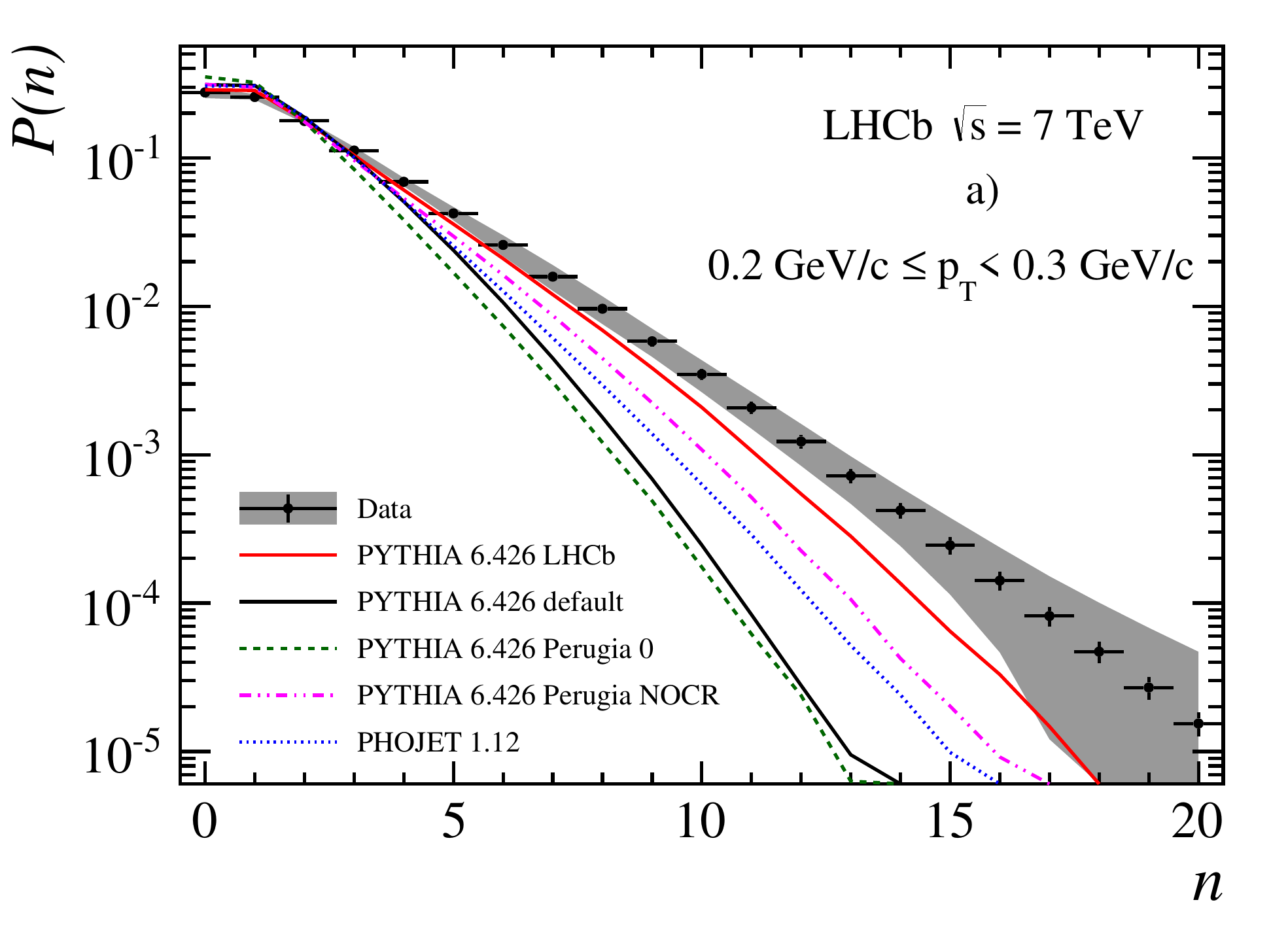}}
    \subfigure{\includegraphics*[width=0.48\textwidth]{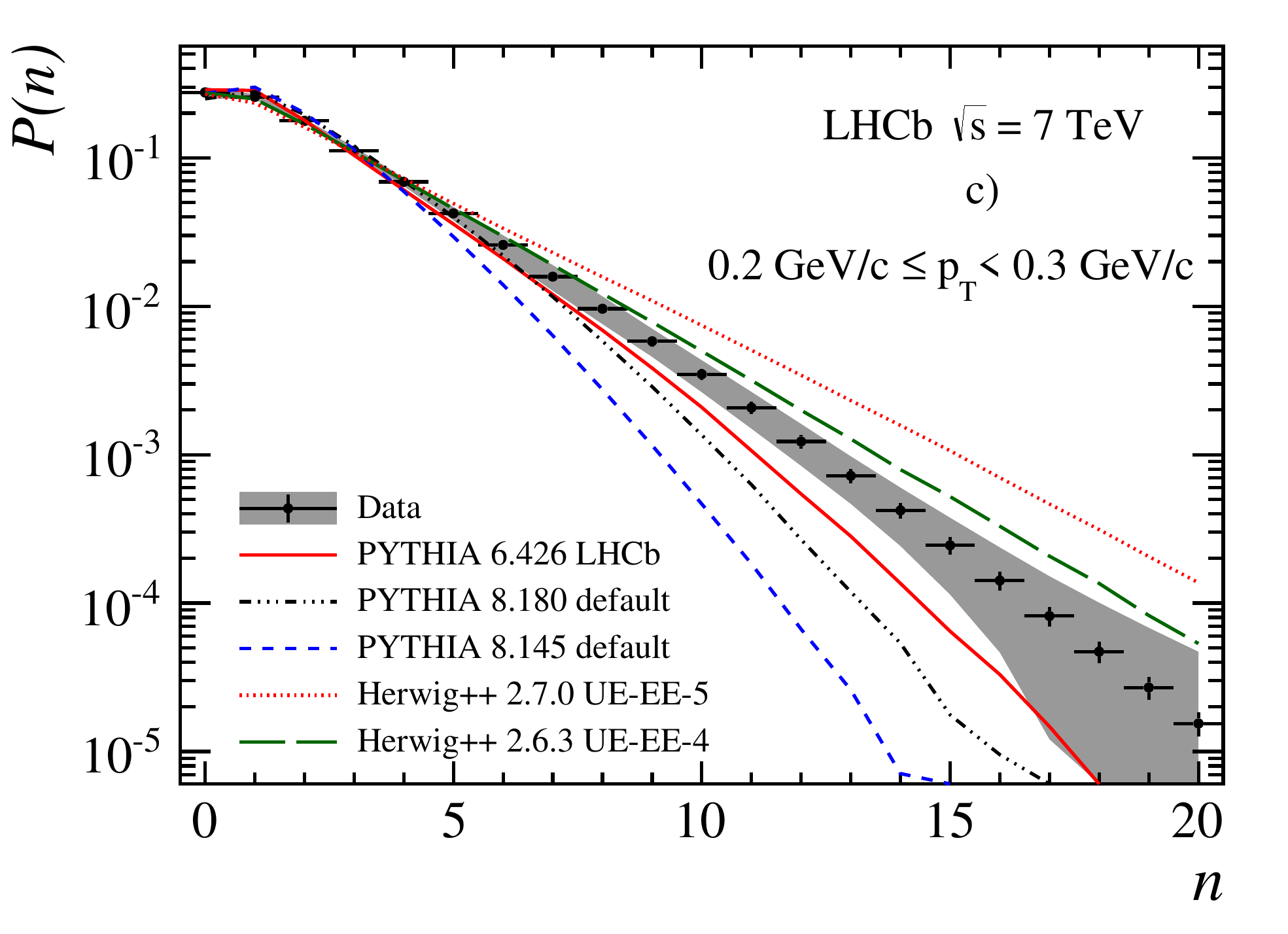}}
    \subfigure{\includegraphics*[width=0.48\textwidth]{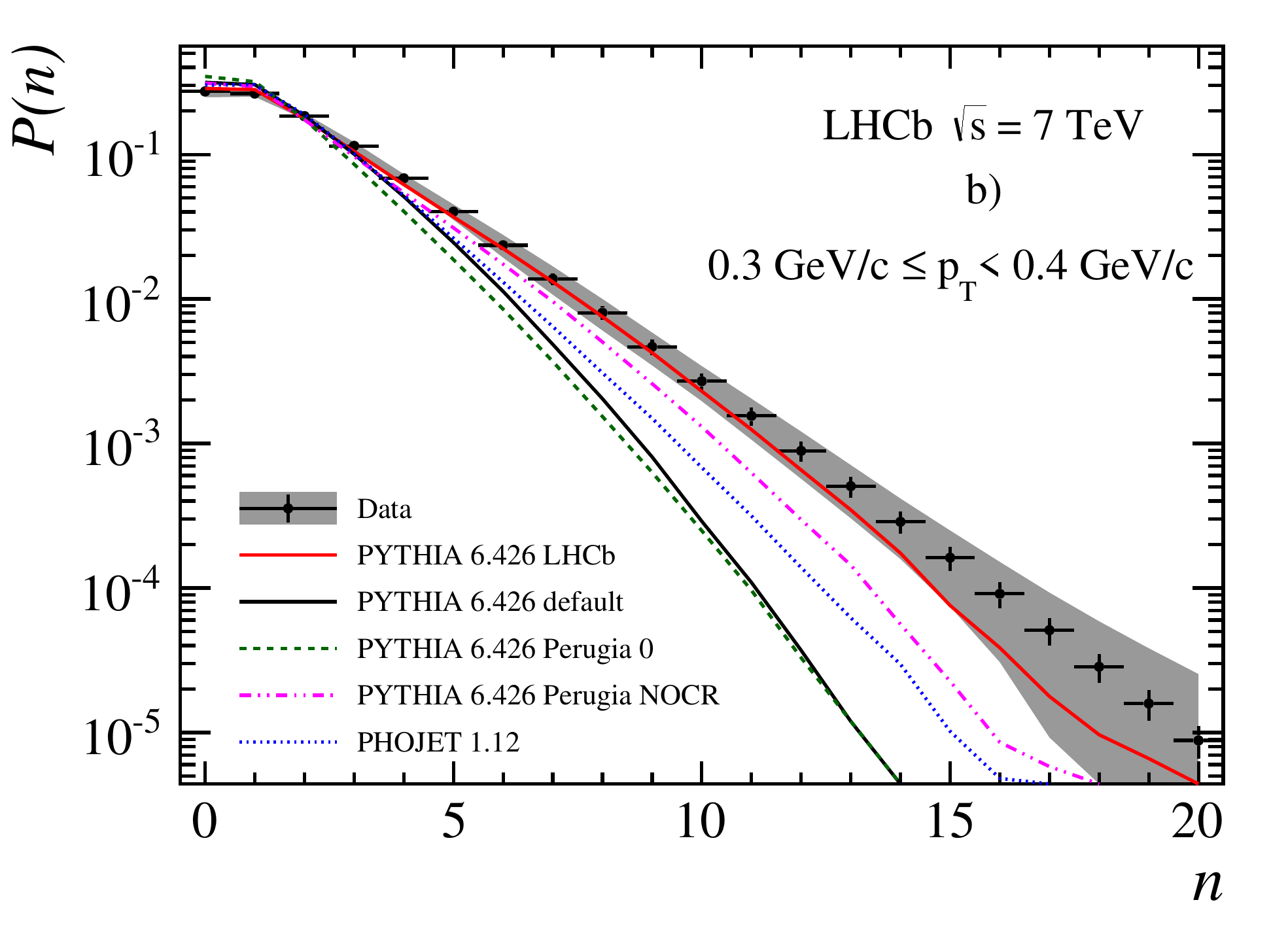}}
    \subfigure{\includegraphics*[width=0.48\textwidth]{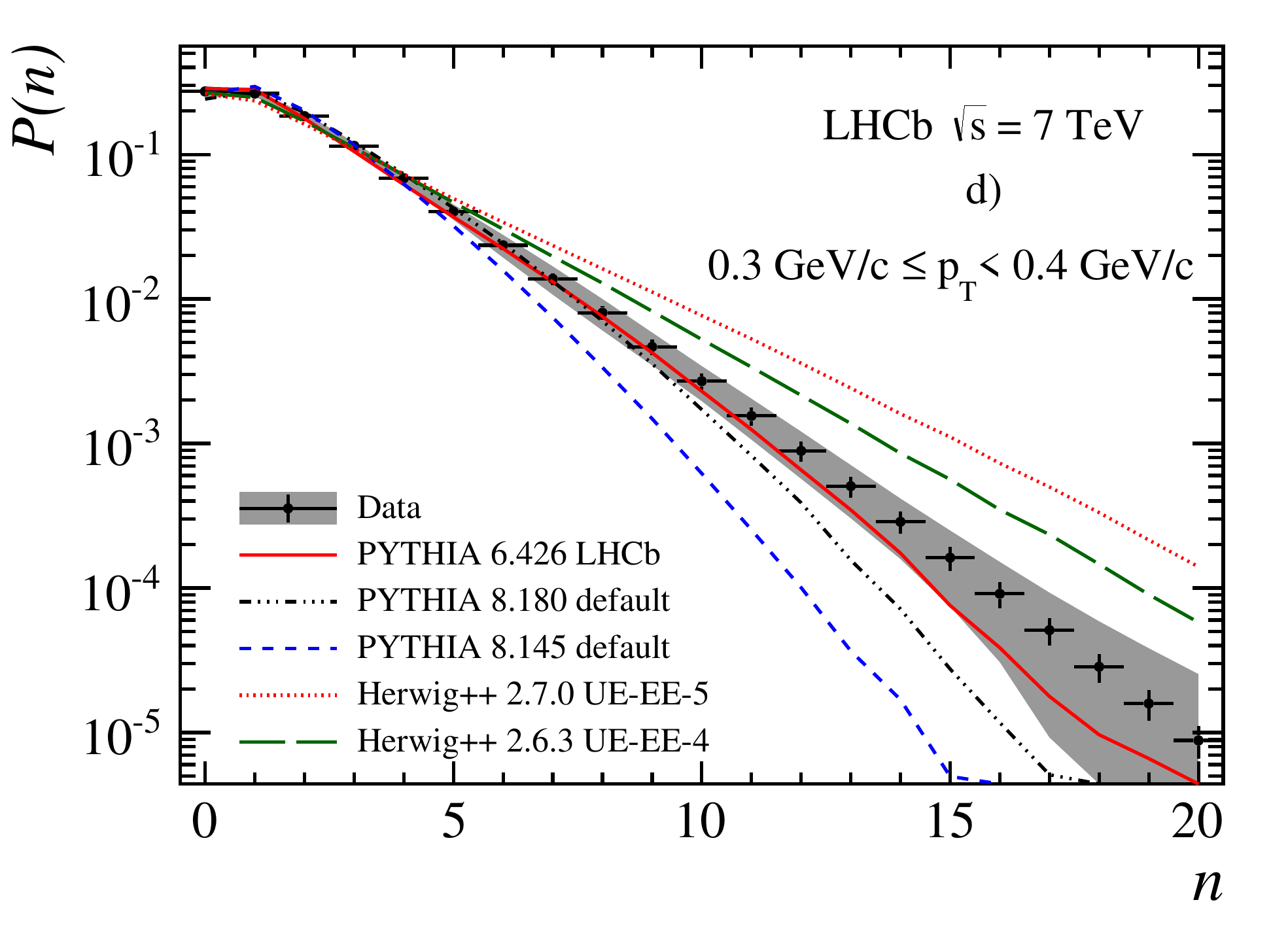}}
  \end{center}
\caption{\small Observed charged particle multiplicity distribution in different \pt bins. 
Error bars represent the statistical uncertainty, the error bands show the combined statistical and systematic uncertainties.
The data are compared to Monte Carlo predictions, (a,b) \pythia 6 and \phojet, (c,d) \pythia 8 and \herwig. 
All plots show predictions of the \lhcb tune of \pythia 6, which is used in the analysis.}
\label{fig:BinnedMult_Pt1}%
\end{figure}
\begin{figure}[h]
  \begin{center}
    \subfigure{\includegraphics*[width=0.48\textwidth]{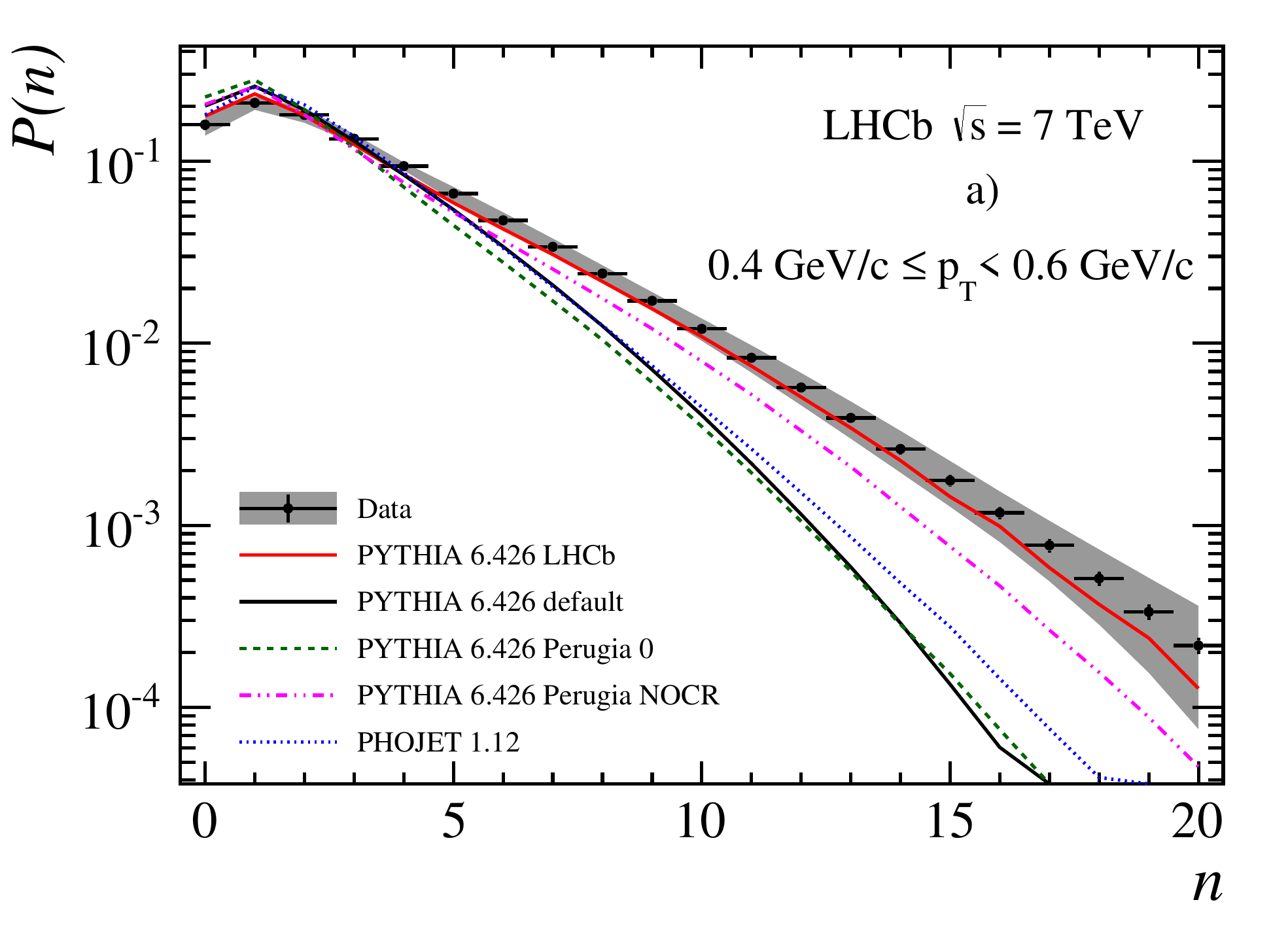}}
    \subfigure{\includegraphics*[width=0.48\textwidth]{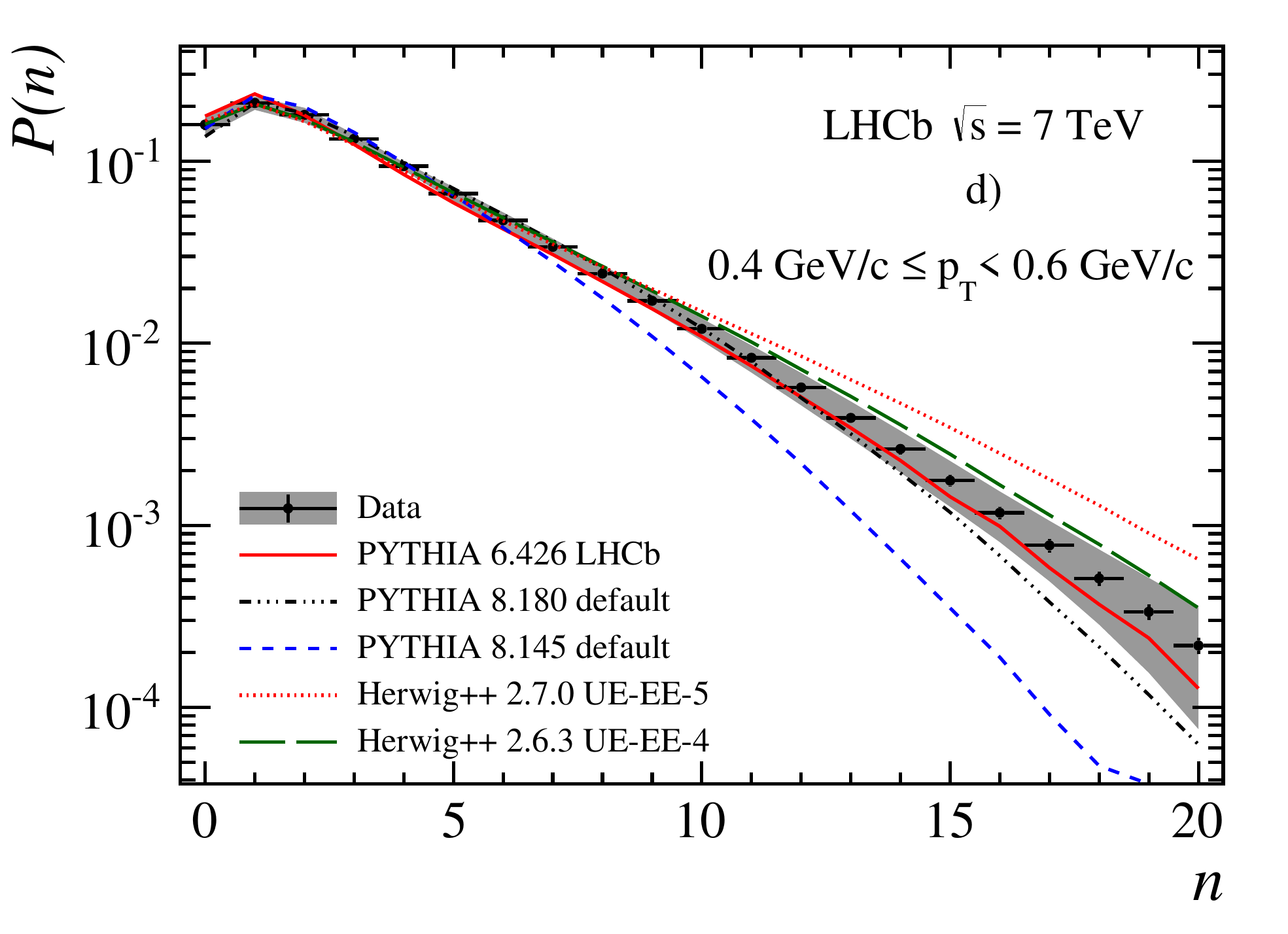}}
    \subfigure{\includegraphics*[width=0.48\textwidth]{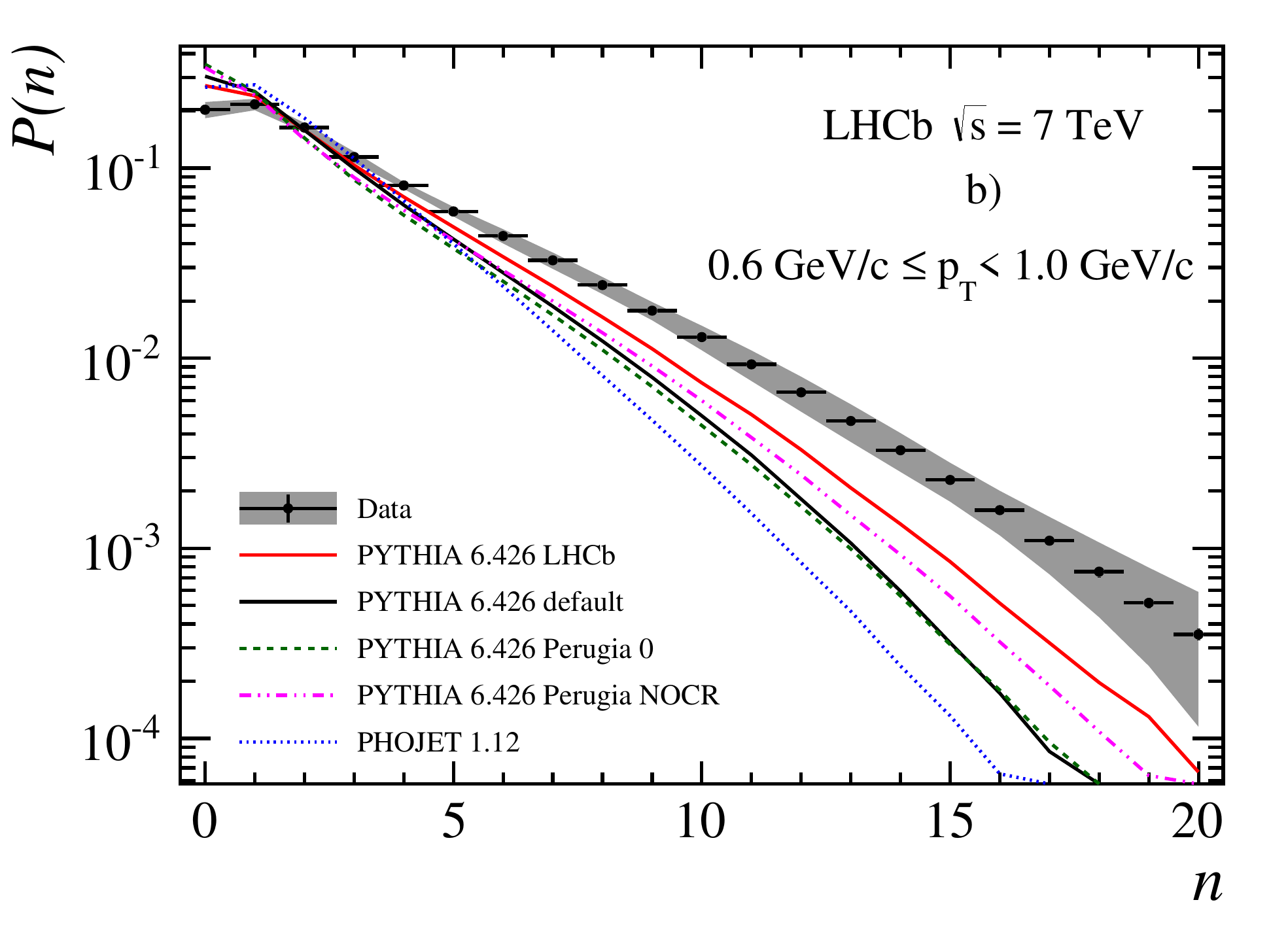}}
    \subfigure{\includegraphics*[width=0.48\textwidth]{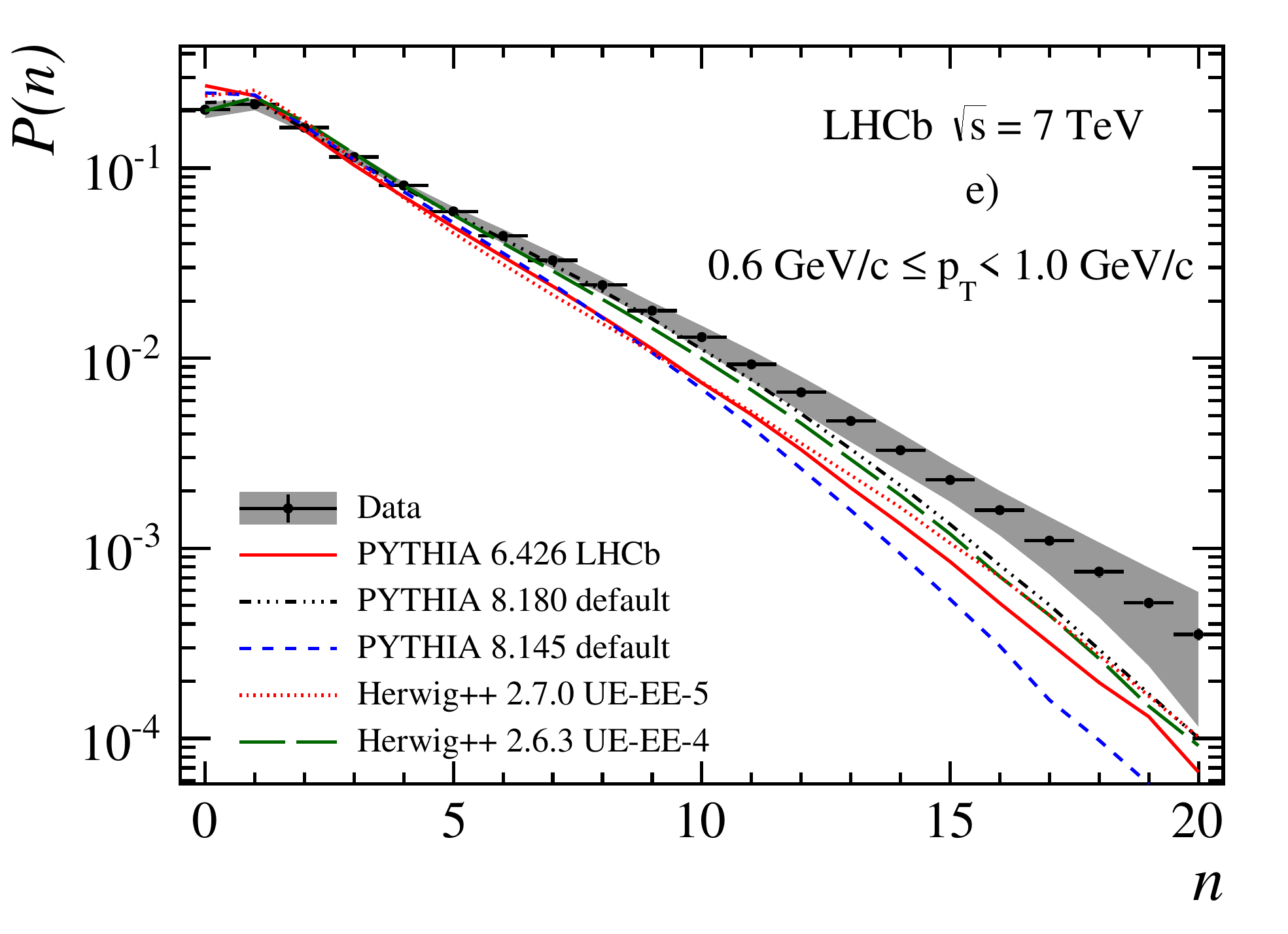}}
    \subfigure{\includegraphics*[width=0.48\textwidth]{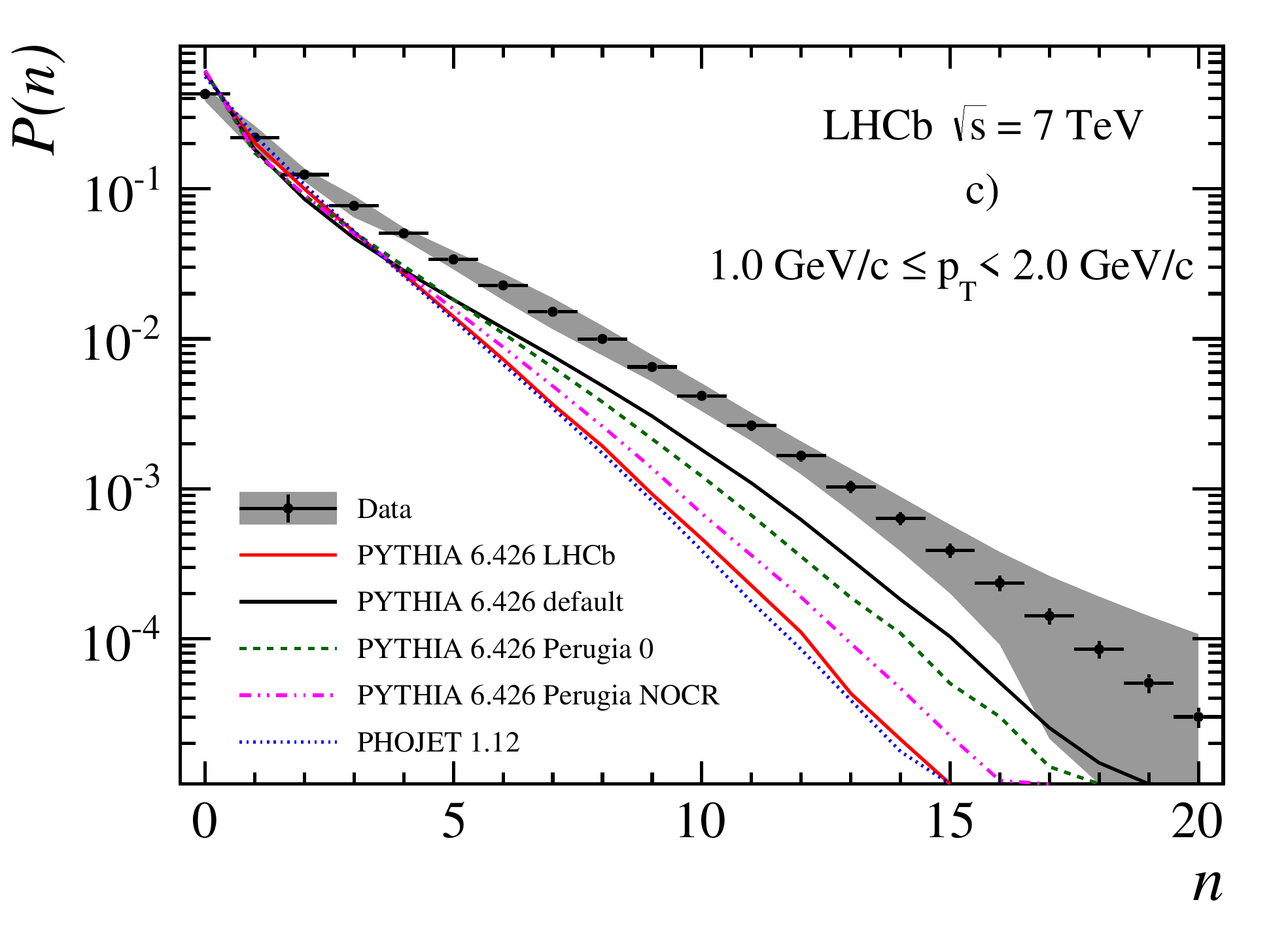}}
    \subfigure{\includegraphics*[width=0.48\textwidth]{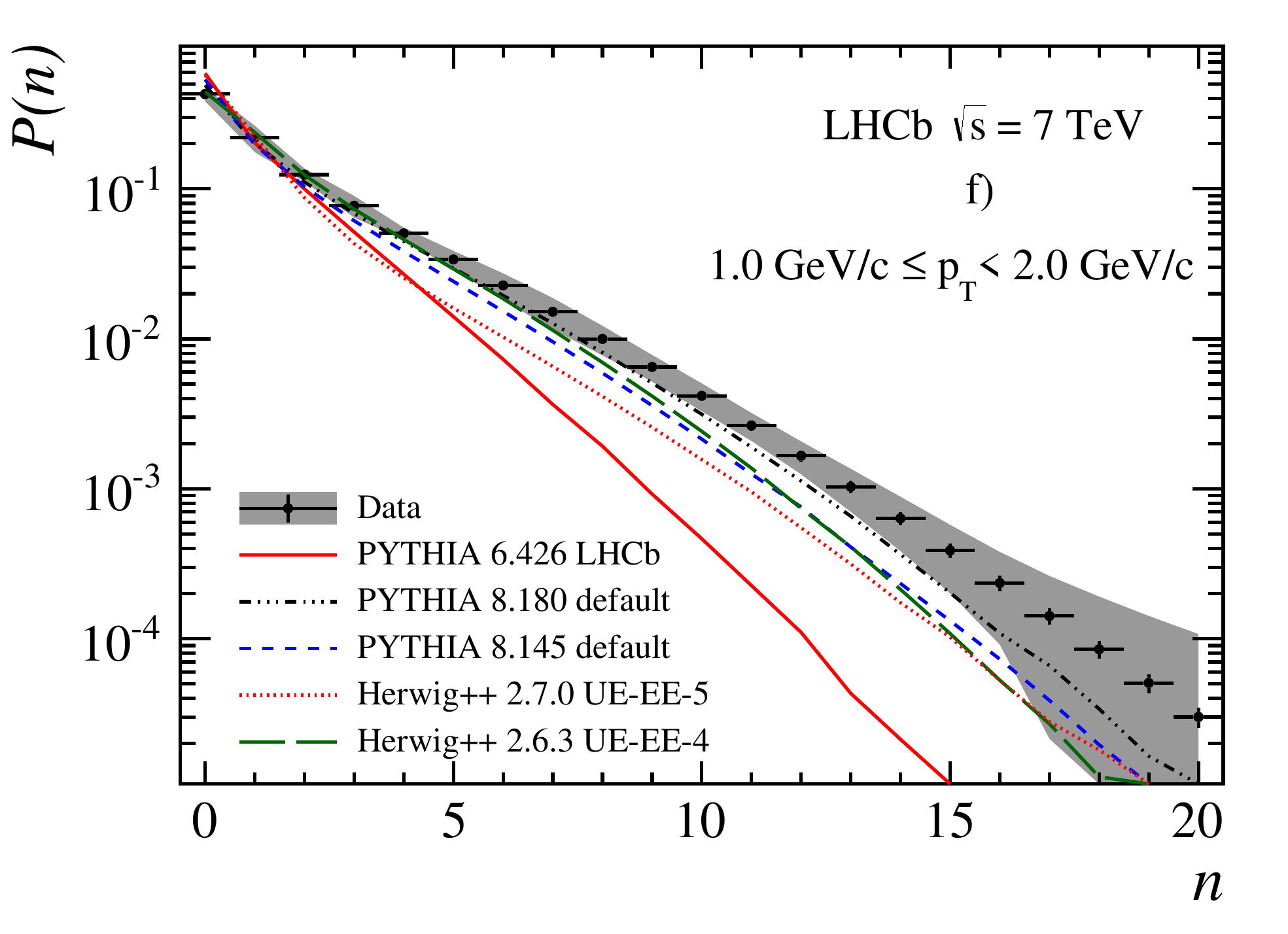}}
  \end{center}
\caption{\small Observed charged particle multiplicity distribution in different \pt bins.
Error bars represent the statistical uncertainty, the error bands show the combined statistical and systematic uncertainties.
The data are compared to Monte Carlo predictions, (a-c) \pythia 6 and \phojet, (d-f) \pythia 8 and \herwig. 
All plots show predictions of the \lhcb tune of \pythia 6, which is used in the analysis.}
\label{fig:BinnedMult_Pt2}%
\end{figure}

Charged particle multiplicities for bins of transverse momentum are shown in Figs.~\ref{fig:BinnedMult_Pt1} and \ref{fig:BinnedMult_Pt2}. 
The LHCb tune describes the data better than the other tunes. 
It is interesting to note that at large transverse momenta, where the discrepancies are most prominent, \pythia 6.426 in the default configuration matches the shape of the distribution. 
\pythia 8 in the recent configuration shows a reasonably good agreement to the measurement in the mid- and high-\pt range, where also the \herwig generator describes the data. 
Predictions using the UE-4 tune are closer to the measurement than using the UE-5 tune. Towards larger \pt, \herwig predictions underestimate the amount of particles while the \pythia 8 prediction is slightly better. 
\pythia 8 underestimates the data towards lower \pt, while \herwig overestimates it. 

The mean value and the root-mean-square deviation for the multiplicity distributions in $\eta$ and \pt bins are tabulated in the Appendix.

%% file: summary.tex

\clearpage

\section{Summary}
\label{sec:Summary}

The charged particle multiplicities and the mean particle densities are measured in inclusive $pp$ interactions at a centre-of-mass energy of $\sqrt{s}=7\;$TeV with the \lhcb detector. 
The measurement is performed in the kinematic range $\ptot>2\gevc$, $\pt>0.2\gevc$ and $2.0<\eta<4.8$, in which at least one charged particle per event is required. 
By using the full spectrometer information, it is possible to extend the previous \lhcb results~\cite{LHCb-PAPER-2011-011} to include momentum dependent measurements. 
The comparison of data with predictions from several Monte Carlo event generators shows that predictions from recent generators, tuned to \lhc measurements in the central rapidity region, are in better agreement than predictions from older generators. 
While the phenomenology in some kinematic regions is well described by recent \pythia and \herwig simulations, the data in the higher \pt and small $\eta$ ranges of the probed kinematic region are still underestimated. 
None of the event generators considered are able to describe the entire range of measurements. 

%% file: acknowledgements.tex
\section*{Acknowledgements}

\noindent We express our gratitude to our colleagues in the CERN
accelerator departments for the excellent performance of the LHC. We
thank the technical and administrative staff at the LHCb
institutes. We acknowledge support from CERN and from the national
agencies: CAPES, CNPq, FAPERJ and FINEP (Brazil); NSFC (China);
CNRS/IN2P3 and Region Auvergne (France); BMBF, DFG, HGF and MPG
(Germany); SFI (Ireland); INFN (Italy); FOM and NWO (The Netherlands);
SCSR (Poland); MEN/IFA (Romania); MinES, Rosatom, RFBR and NRC
``Kurchatov Institute'' (Russia); MinECo, XuntaGal and GENCAT (Spain);
SNSF and SER (Switzerland); NAS Ukraine (Ukraine); STFC (United
Kingdom); NSF (USA). We also acknowledge the support received from the
ERC under FP7. The Tier1 computing centres are supported by IN2P3
(France), KIT and BMBF (Germany), INFN (Italy), NWO and SURF (The
Netherlands), PIC (Spain), GridPP (United Kingdom).
We are indebted to the communities behind the multiple open source software packages we depend on.
We are also thankful for the computing resources and the access to software R\&D tools provided by Yandex LLC (Russia).

%% file: appendix.tex

\clearpage

{\noindent\bf\Large Appendix}

\appendix

\begin{figure}[!h]
  \begin{center}
    \includegraphics*[width=0.70\textwidth]{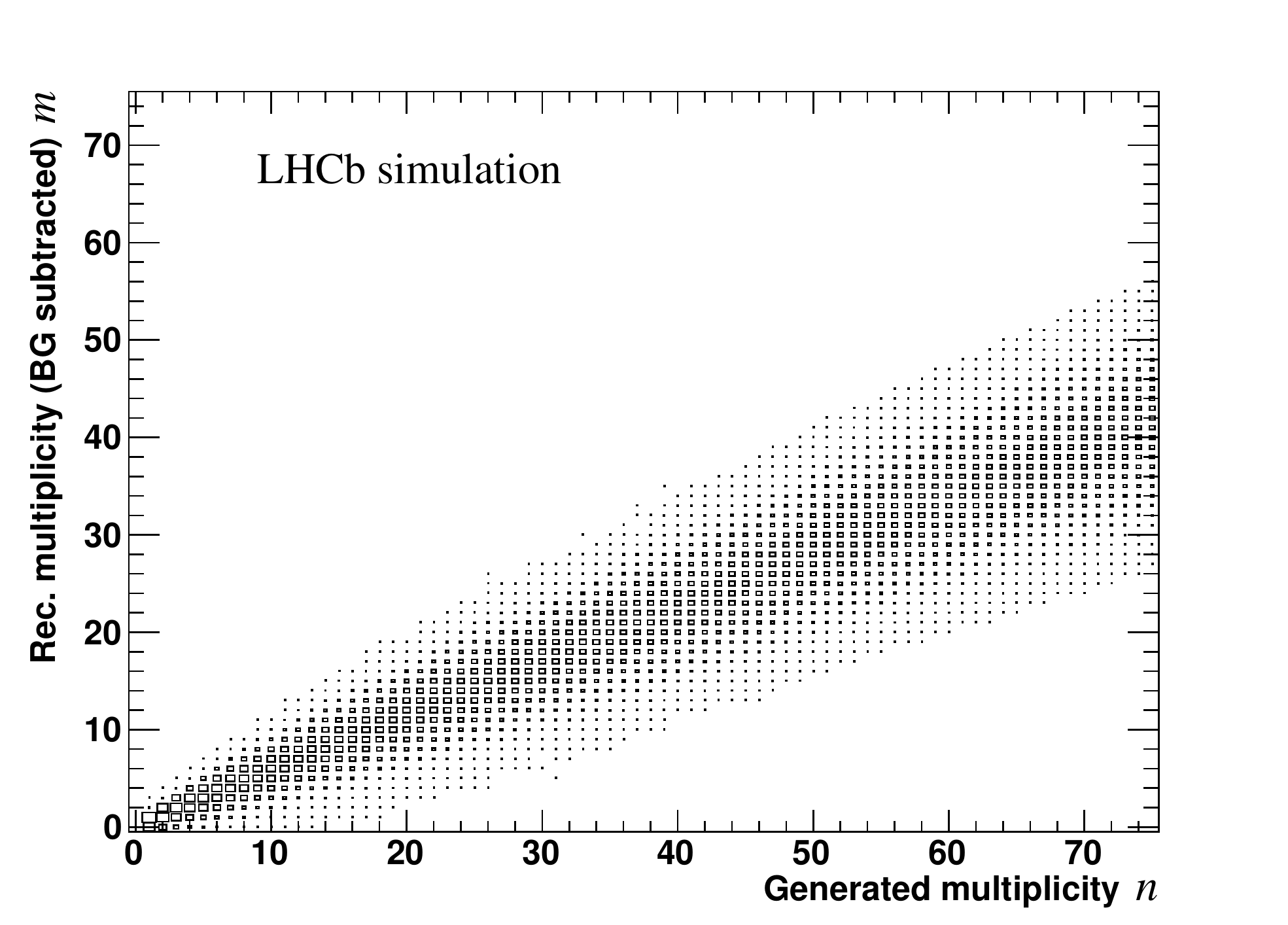}
  \end{center}
\caption{\small Example of the parametrized detector response matrix in the full kinematic range. The matrix is obtained from fully simulated events showing the relation between the true charged particle multiplicity and the reconstructed and background subtracted track multiplicity.}
\label{fig:ResponseMatrix}%
\end{figure}

\begin{table}[h]
\centering
\renewcommand{\arraystretch}{1.3} 
\begin{tabular}{c|c}
 Pseudorapidity range & $dn/d\eta $\\
  \hline
  $2.0 \leq \eta < 2.2$ & $3.600\pm0.048\pm0.463$ \\
  
  $2.2 \leq \eta < 2.4$ & $4.032\pm0.050\pm0.460$ \\
  
  $2.4 \leq \eta < 2.6$ & $4.428\pm0.055\pm0.367$ \\
  
  $2.6 \leq \eta < 2.8$ & $4.754\pm0.056\pm0.277$ \\
  
  $2.8 \leq \eta < 3.0$ & $4.943\pm0.057\pm0.285$ \\
  
  $3.0 \leq \eta < 3.2$ & $4.977\pm0.055\pm0.267$ \\
  
  $3.2 \leq \eta < 3.4$ & $4.734\pm0.052\pm0.213$ \\
  
  $3.4 \leq \eta < 3.6$ & $4.500\pm0.050\pm0.207$ \\
  
  $3.6 \leq \eta < 3.8$ & $4.267\pm0.049\pm0.200$ \\
  
  $3.8 \leq \eta < 4.0$ & $4.026\pm0.047\pm0.194$ \\
  
  $4.0 \leq \eta < 4.2$ & $3.845\pm0.046\pm0.186$ \\
  
  $4.2 \leq \eta < 4.4$ & $3.613\pm0.047\pm0.263$ \\

  $4.4 \leq \eta < 4.6$ & $3.358\pm0.043\pm0.179$ \\

  $4.6 \leq \eta < 4.8$ & $3.281\pm0.045\pm0.174$ \\
  \hline
\end{tabular} 
\caption{Charged particle density as a function of pseudorapidity. The first quoted uncertainty is statistical and the second systematic.}
\label{tab:DensityEta}
\end{table}

\begin{table}[h]
\centering
\renewcommand{\arraystretch}{1.3} 
\begin{tabular}{c|c}
 Transverse momentum range [\gevc] &  $dn/d\pt \;[0.1\gevc]^{-1}$ \\ 
  
  \hline
  $0.20 \leq \pt < 0.30$ & $1.908\pm0.024\pm0.116$ \\

  $0.30 \leq \pt < 0.40$ & $1.866\pm0.026\pm0.099$ \\

  $0.40 \leq \pt < 0.50$ & $1.678\pm0.022\pm0.093$ \\

  $0.50 \leq \pt < 0.60$ & $1.347\pm0.009\pm0.092$ \\

  $0.60 \leq \pt < 0.70$ & $1.082\pm0.007\pm0.091$ \\

  $0.70 \leq \pt < 0.80$ & $0.817\pm0.006\pm0.064$ \\

  $0.80 \leq \pt < 0.90$ & $0.617\pm0.006\pm0.042$ \\

  $0.90 \leq \pt < 1.00$ & $0.481\pm0.005\pm0.044$ \\

  $1.00 \leq \pt < 1.10$ & $0.366\pm0.005\pm0.019$ \\

  $1.10 \leq \pt < 1.20$ & $0.290\pm0.004\pm0.015$ \\
  
  $1.20 \leq \pt < 1.30$ & $0.228\pm0.004\pm0.012$ \\

  $1.30 \leq \pt < 1.40$ & $0.180\pm0.004\pm0.009$ \\
  
  $1.40 \leq \pt < 1.50$ & $0.144\pm0.003\pm0.007$ \\
  
  $1.50 \leq \pt < 1.60$ & $0.113\pm0.002\pm0.007$ \\
  
  $1.60 \leq \pt < 1.70$ & $0.092\pm0.002\pm0.006$ \\
  
  $1.70 \leq \pt < 1.80$ & $0.075\pm0.001\pm0.005$ \\
  
  $1.80 \leq \pt < 1.90$ & $0.061\pm0.001\pm0.004$ \\
  
  $1.90 \leq \pt < 2.00$ & $0.053\pm0.001\pm0.003$ \\
  \hline
\end{tabular} 
\caption{Charged particle density as a function of transverse momentum. The first quoted uncertainty is statistical and the second systematic.}
\label{tab:DensityPt}
\end{table}

\begin{table}[h]
\centering
\begin{tabular}{c|c|c}
 Pseudorapidity range & Mean value  &  Root-mean-square \\
  \hline
    $2.0 \leq \eta < 2.5$ & $2.010 \pm 0.002 \pm 0.118$         & $2.460 \pm 0.002 \pm 0.115$ \\  

    $2.5 \leq \eta < 3.0$ & $2.424 \pm 0.002 \pm 0.097$         & $2.736 \pm 0.002 \pm 0.094$ \\  

    $3.0 \leq \eta < 3.5$ & $2.409 \pm 0.002 \pm 0.100$         & $2.668 \pm 0.002 \pm 0.113$ \\  

    $3.5 \leq \eta < 4.0$ & $2.121 \pm 0.002 \pm 0.087$         & $2.396 \pm 0.001 \pm 0.117$ \\  

    $4.0 \leq \eta < 4.5$ & $1.852 \pm 0.002 \pm 0.069$         & $2.093 \pm 0.001 \pm 0.073$ \\  
   \hline
   \end{tabular} 
\caption{Truncated mean value and root-mean-square deviation for charged particle multiplicities in different $\eta$-bins. The range is from 0 to 20 particles. The first quoted uncertainty is statistical and the second systematic.}
\label{tab:MultiplicityEta}
\end{table}

\begin{table}[h]
\centering
\begin{tabular}{c|c|c}
 Transverse momentum range [\gevc] & Mean value & Root-mean-square  \\
  \hline
    $0.2 \leq \pt < 0.3$ & $1.928 \pm 0.002 \pm 0.073$  & $2.083 \pm 0.001 \pm 0.067$ \\  

    $0.3 \leq \pt < 0.4$ & $1.865 \pm 0.002 \pm 0.065$  & $1.971 \pm 0.001 \pm 0.050$ \\  

    $0.4 \leq \pt < 0.6$ & $2.988 \pm 0.002 \pm 0.098$  & $2.855 \pm 0.002 \pm 0.069$ \\  

    $0.6 \leq \pt < 1.0$ & $2.881 \pm 0.003 \pm 0.103$  & $3.029 \pm 0.002 \pm 0.090$ \\  

    $1.0 \leq \pt < 2.0$ & $1.580 \pm 0.002 \pm 0.096$  & $2.195 \pm 0.001 \pm 0.093$ \\  
   \hline
   \end{tabular} 
\caption{Truncated mean value and root-mean-square deviation for charged particle multiplicities in different \pt-bins. The range is from 0 to 20 particles. The first quoted uncertainty is statistical and the second systematic.}
\label{tab:MultiplicityPt}
\end{table}

\clearpage